\documentclass[aps,twocolumn,showpacs,preprintnumbers,prl,10pt]{revtex4-1}
\usepackage[utf8]{inputenc}
\usepackage{graphicx,amssymb,amsmath,amsthm,amsfonts,epsfig,times,natbib}
\usepackage{subfigure}
\usepackage[usenames,dvipsnames]{color}
\usepackage{epstopdf}
\usepackage{amsmath,amssymb}
\usepackage{tensor}
\usepackage{mathtools}
\usepackage{amsbsy}
\usepackage{bm,url}

\usepackage[scaled]{beramono}
\usepackage[T1]{fontenc}
\usepackage[linktocpage]{hyperref}

\definecolor{oxfordblue}{rgb}{0.0, 0.13, 0.28}
\definecolor{burgundy}{rgb}{0.5, 0.0, 0.13}
\definecolor{darkolivegreen}{rgb}{0.33, 0.42, 0.18}
\definecolor{darkblue}{rgb}{0,0,0.5}
\definecolor{richcarmine}{rgb}{0.84, 0.0, 0.25}
\definecolor{darkblue}{rgb}{0,0,0.5}
\definecolor{venetianred}{rgb}{0.78, 0.03, 0.08}
\definecolor{skobeloff}{rgb}{0.0, 0.48, 0.45}
\hypersetup{colorlinks=true, citecolor=darkblue, linkcolor=darkblue,
urlcolor = darkblue}

\newcommand{\ben}{\begin{enumerate}}
\newcommand{\een}{\end{enumerate}}

\def\be{\begin{equation}}
\def\ee{\end{equation}}
\def\bea{\begin{eqnarray}}
\def\eea{\end{eqnarray}}

\newcommand{\beq}{\begin{eqnarray}}
\newcommand{\eeq}{\end{eqnarray}} 
\newcommand{\ba}{\begin{align}}
\newcommand{\ea}{\end{align}}

\begin{document}

\title{Splitting the Gravitational Atom: \\ Instabilities of Black Holes with Synchronized or Resonant Hair}

\author{Jordan Nicoules}
\affiliation{Departamento de Matem\'atica da Universidade de Aveiro and CIDMA,
Campus de Santiago, 3810-193 Aveiro, Portugal}

\author{José Ferreira}
\affiliation{Departamento de Matem\'atica da Universidade de Aveiro and CIDMA,
Campus de Santiago, 3810-193 Aveiro, Portugal}

\author{Carlos A. R. Herdeiro}
\affiliation{Departamento de Matem\'atica da Universidade de Aveiro and CIDMA,
Campus de Santiago, 3810-193 Aveiro, Portugal}

\author{Eugen Radu}
\affiliation{Departamento de Matem\'atica da Universidade de Aveiro and CIDMA,
Campus de Santiago, 3810-193 Aveiro, Portugal}

\author{Miguel Zilhão}
\affiliation{Departamento de Matem\'atica da Universidade de Aveiro and CIDMA,
Campus de Santiago, 3810-193 Aveiro, Portugal}


\date{October 2025.
Published: March 2026.
}

\begin{abstract}
Black holes (BHs) with synchronized bosonic hair challenge the Kerr paradigm, linking superradiance from ultralight fields -- creating \textit{gravitational atoms} -- to bosonic stars across parameter space. In the ``very hairy'' regime, where a small horizon lies inside a bosonic star containing most of the energy, they deviate sharply from Kerr, but their dynamics remain unexplored. We show that for such solutions the horizon gets naturally ejected from the center of its scalar environment and observe a similar dynamics in a cousin model of BHs with resonant scalar hair, albeit with a different fate. This dynamical splitting is likely to be generic for sufficiently hairy BHs in the broader class of models with synchronized or resonant hair, but possible exceptions may exist.
\end{abstract}


\maketitle

{\bf Introduction.}
Black holes (BHs) ``have no \textit{hair}''~\cite{Ruffini:1971} is a community dictum that reflects the surprising simplicity of BHs in electrovacuum general relativity (GR). However, beyond this paradigmatic model, BHs \textit{can} have hair~\cite{Herdeiro:2015waa,Volkov:2016ehx}: more macroscopic degrees of freedom not associated with gauge symmetries. Then, the key question becomes whether such hairy models are viable, both theoretically and phenomenologically, as alternatives to the Kerr hypothesis~\cite{Cardoso:2016ryw,Herdeiro:2022yle}. 

Some non-Kerr -- including hairy -- BHs~\cite{Barack:2018yly} are model-specific; some rely on general mechanisms. An example of the latter is the family of \textit{BHs with synchronized (bosonic) hair} (BHsSH)~\cite{Herdeiro:2014goa}. 
Dynamical synchronization is a ubiquitous phenomenon in physical and biological systems~\cite{strogatz2003sync}; a familiar example is the Moon always showing the same face toward Earth. In this spirit, BHsSH have a \textit{dynamical} rationale~\cite{Herdeiro:2025ubg}: simulations show that they form from superradiance~\cite{East:2017ovw,Herdeiro:2017phl}, yielding a sort of \textit{gravitational atom}~\footnote{The term gravitational atom has been often used for describing quasibound states around Kerr BHs~\cite{Baumann:2019eav}; here we use it for true bound states, including those with backreaction, forming ``hairy'' BHs.}, and perturbative studies that they can be long-lived, possibly for cosmological timescales~\cite{Ganchev:2017uuo,Degollado:2018ypf}. But this dynamical picture is valid for small amounts of hair only~\cite{Herdeiro:2021znw}.  

Dynamics is, in fact, a key viability discriminator: whether the solution can form and be sufficiently long-lived. Then, it could be a plausible player in (astro)physical processes, justifying a comparison with state-of-the-art  data~\cite{Abbott:2016blz,Akiyama:2019cqa}.
But dynamical studies of non-Kerr models can be challenging, especially for spinning BHs, the most common astrophysical players. In fact, the Kerr metric lends itself to a linear perturbative analysis~\cite{Whiting:1988vc} that is not available for generic non-Kerr models. Nonlinear stability or formation studies are also difficult, relying mostly on numerical relativity, under control for vacuum GR and a handful of other models only. Fortunately, BHsSH arise in the latter class and are therefore within the grasp of these powerful tools, already used for evolving spinning bosonic stars~\cite{Schunck:2003kk,Brito:2015pxa}, the solitonic limit of BHsSH~\cite{Sanchis-Gual:2019ljs}. 

This paper addresses the dynamics of the very hairy BHsSH using numerical relativity, where notable non-Kerr phenomenology arises~\cite{Cunha:2015yba,Ni:2016rhz,Collodel:2021jwi}.
In the simplest example of very hairy BHsSH we show that the horizon and the hair tend to split, and a similar behaviour is found in a cousin model. 
This result rules out the physical viability of BHs with large hair in these models, corroborating the lack of formation mechanisms in this region of the parameter space~\cite{East:2017ovw,Herdeiro:2021znw,Sanchis-Gual:2020mzb}. 
%
{\bf Equilibrium BHs.}
The simplest model of BHsSH occurs for a massive, complex scalar field $\Phi$ minimally coupled to GR: $ \mathcal{S} = \int d^4x \sqrt{-g}\left[{R}/(16\pi G)+\mathcal{L}_{\rm m}\right]$, where $(c=1=\hbar)$
\begin{equation}
  \mathcal{L}_{\rm m} = - \nabla_\alpha\Phi^*\nabla^\alpha\Phi - \mu^2\left|\Phi\right|^2  \ . 
  \label{action}
\end{equation}  
$R$ and $G$ are the Ricci scalar and Newton's constant, $g$ and $\mu$ are the metric determinant and scalar field mass, and $*$ denotes the complex conjugate. BHsSH are numerically computed with the \textit{Ansatz}
$ds^2 = -e^{F_0}Ndt^2 + e^{2F_1}\left({dr^2}/{N} + r^2d\theta^2 \right)  
	+ e^{2F_2}r^2\sin^2\theta \left( d\varphi - Wdt \right)^2$ and $N\equiv 1-r_H/r$,
where  
$F_i,W$, $i=1,2,3$ are functions of the spheroidal coordinates $r,\theta$; $r_H$ is the horizon radial coordinate and $W(r_H)=\text{const}=\Omega_H$  is the  horizon angular velocity.
The scalar field \textit{Ansatz} is 
$\Phi = e^{ i ( m\varphi -\omega t)}\phi(r,\theta)$, where the scalar field frequency $\omega>0$  and $m\in \mathbb{Z^+}$. We focus on radially nodeless solutions [$\phi(r,\theta=\pi/2)$ has no zeros] with $m=1$, corresponding to the fundamental spinning BHsSH~\cite{Herdeiro:2019mbz}. The synchronization
condition states that $\Omega_H=\omega/m$.

The space of solutions is described in \cite{Herdeiro:2014goa,Herdeiro:2015gia}. Two limiting regimes are (i) for vanishing scalar field $(\Phi\rightarrow 0)$, a subset of Kerr BHs are obtained, computed using a test scalar field on Kerr~\cite{Hod:2012px} -- nearby, one finds almost bald BHsSH; (ii) for vanishing BH size (implying $r_H\rightarrow 0$), spinning boson stars are obtained~\cite{Schunck:1996he,Yoshida:1997qf}. These have a \textit{toroidal} energy distribution. Nearby one finds very hairy BHs, whose dynamics can be understood via a test particle approximation on the boson stars geometry. BHsSH have global (resp. horizon) mass and angular momentum $M$ and $J$ ($M_H$ and $J_H)$, respectively~\cite{Herdeiro:2015gia}.

The (perturbative) dynamics of almost bald BHsSH uncovered superradiant instabilities~\cite{Ganchev:2017uuo}, anticipated in~\cite{Herdeiro:2014jaa}. These could be very long-lived~\cite{Degollado:2018ypf}. Here we shall focus on the nonlinear dynamics of very hairy BHs, unexplored so far. Some physical quantities of three illustrative solutions considered below are presented in the following table (henceforth using $\mu=1$ and $G=1$ unless explicitly stated):

\begin{table}[h]
    \begin{tabular}{c|cc|cccc}
         BHsSH &  $\omega$ & $r_H$ & $M$ & $J$ & $M_H/M$ & $J_H/J$  \\\hline
         \textbf{A} &  0.998 & 0.18 & 0.1277 & 0.0347 & 0.7550 & 0.0998 \\
         \textbf{B} &  0.83 & 0.10 & 1.2170 & 1.2021 & 0.0419 & 0.0014  \\
         \textbf{C} &  0.90 & 0.20 & 1.0105 & 0.9119 & 0.1175 & 0.0122 
    \end{tabular}
    \label{tab:HairyBH_configs}
\end{table}

\begin{figure}[h!]
\begin{center}
\includegraphics[width=0.45\textwidth]{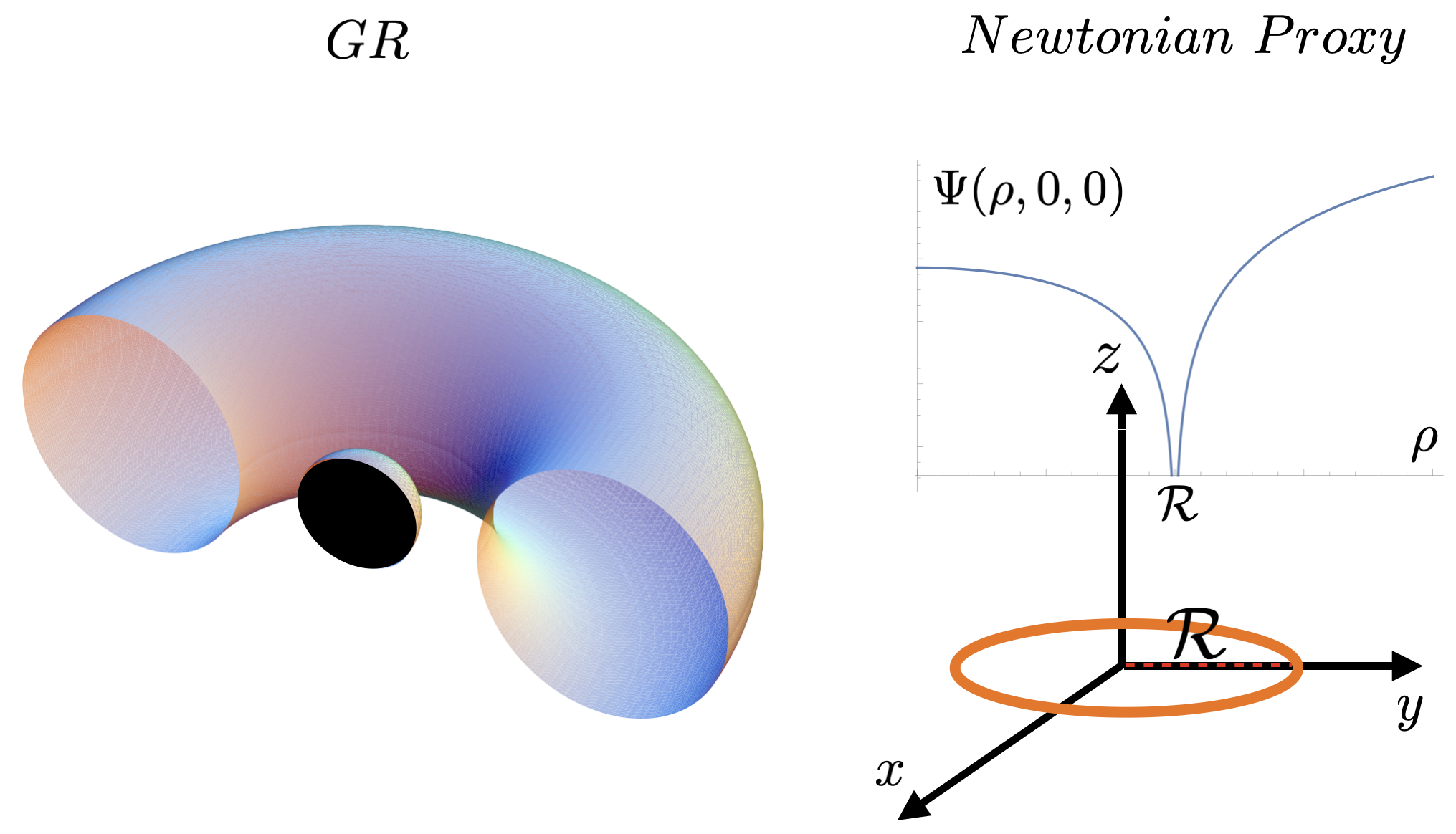}
\caption{\small Left: BHsSH in the simplest scalar model -- a horizon inside a toroidal boson star. Right: a thin ring of constant mass density (bottom) and its Newtonian potential (top).}
\label{fig1}
\end{center}
\end{figure}

{\bf A toy model.}
Since BHsSH interpolate between Kerr BHs and spinning boson stars they may be regarded as an equilibrium non-linear bound state between these limiting configurations -- Fig.~\ref{fig1} (left). Is this equilibrium \textit{stable}?

\vspace{-0.3cm}

It turns out that a small horizon, modeled as a test particle, lurking at the center of a more massive boson star is in an unstable equilibrium~\cite{Delgado:2021jxd} and the system splits when perturbed. This key observation can be corroborated by a Newtonian proxy and a simple computation. Since spinning boson stars are toroidal, consider a thin ring, radius $\mathcal{R}$, and constant mass density $\chi$ in Newtonian gravity -- Fig.~\ref{fig1} (bottom right). In a cylindrical chart, take the ring in the equatorial plane, centered at the origin; its gravitational potential $\Psi$ at a point $(\rho,z,\varphi)$ reads $\Psi(\rho,z,\varphi)=-\int_0^{2\pi}\chi \, \mathcal{R} \,(\rho^2-2\rho\mathcal{R}\cos(\varphi -\tilde{\varphi})+\mathcal{R}^2+z^2)^{-1/2}d\tilde{\varphi}$. Along an equatorial radius, it is represented in Fig.~\ref{fig1} (top right) and near the origin it reads $\Psi(\rho,0,0)\simeq -2\pi \chi-{\pi}\chi\rho^2/(2\mathcal{R}^2)$. Thus, the origin is an unstable equilibrium for radial displacements (not in $z$). Under a small (radial) perturbation, a particle at the origin starts falling towards the ring. 

BHsSH gravitationally dominated by the toroidal scalar environment should, therefore, be mechanically unstable. However, in their GR counterpart, rotational dynamics should trigger an outspiral motion, rather than radial, as confirmed next.

\smallskip
{\bf Non-linear dynamics of BHsSH.}
We use the Einstein equations in the Baumgarte-Shapiro-Shibata-Nakamura form~\cite{PhysRevD.59.024007, PhysRevD.52.5428} and evolve them numerically 
using the \texttt{Einstein Toolkit} infrastructure~\cite{maxwell_rizzo_2025_15520463,Loffler:2011ay,Zilhao:2013hia}.
We use \texttt{Carpet}~\cite{Carpet} for mesh refinement capabilities,
\texttt{AHFinderDirect}~\cite{AHFinderDirect} for tracking apparent horizons and \texttt{QuasiLocalMeasures}~\cite{QuasiLocalMeasures} to compute horizon quantities.
The spacetime metric and matter fields are evolved using the \texttt{LeanBSSNMoL} and \texttt{Scalar}  \texttt{Cactus} \emph{thorns}, as detailed in~\cite{Sperhake:2006cy,Cunha:2017wao,Zilhao:2015tya} and available through the \texttt{Canuda} library~\cite{witek_2023_7791842}. \texttt{Kuibit} is used for output analysis~\cite{Bozzola2021}.
Evolutions are performed only for $z\geq 0$ by imposing $\mathbb{Z}_2$ symmetry.

The previously described BHsSH are used as initial data. The latter are obtained numerically, covering the exterior BH region, $r_H\geq 0$ -- see~\cite{Herdeiro:2015gia}. Thus, we introduce a quasi-isotropic radial coordinate $R$, related to $r$ by
$    r  = R \left(1 + {r_H}/{4R} \right)^2$. 
This allows the initial data to cover the entire radial coordinate $R$: the horizon is initially located at $R=r_H/4$ and the points inside this radius correspond to a copy of $r>r_H$, with a puncture -- identified with the BH location -- at $R=0$. The $(R,\theta,\varphi)$ coordinates are mapped to Cartesian $(x,y,z)$ in the standard way.

\begin{figure}[h!]
    \centering
    \subfigure{\includegraphics[width=0.32\columnwidth]{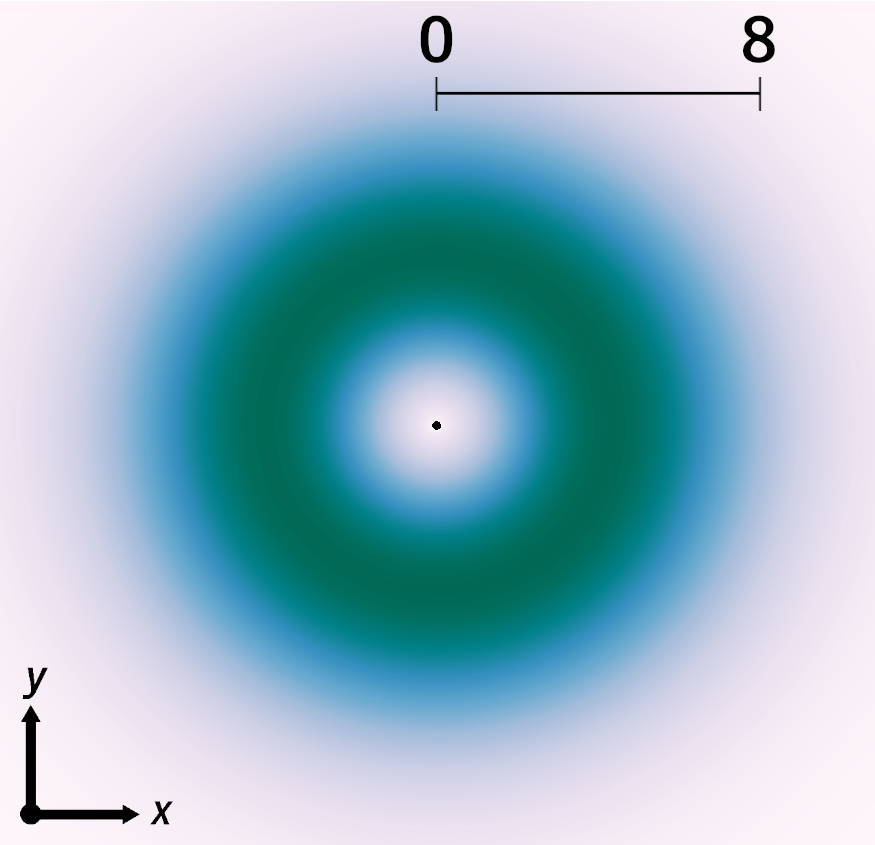}}
    \subfigure{\includegraphics[width=0.32\columnwidth]{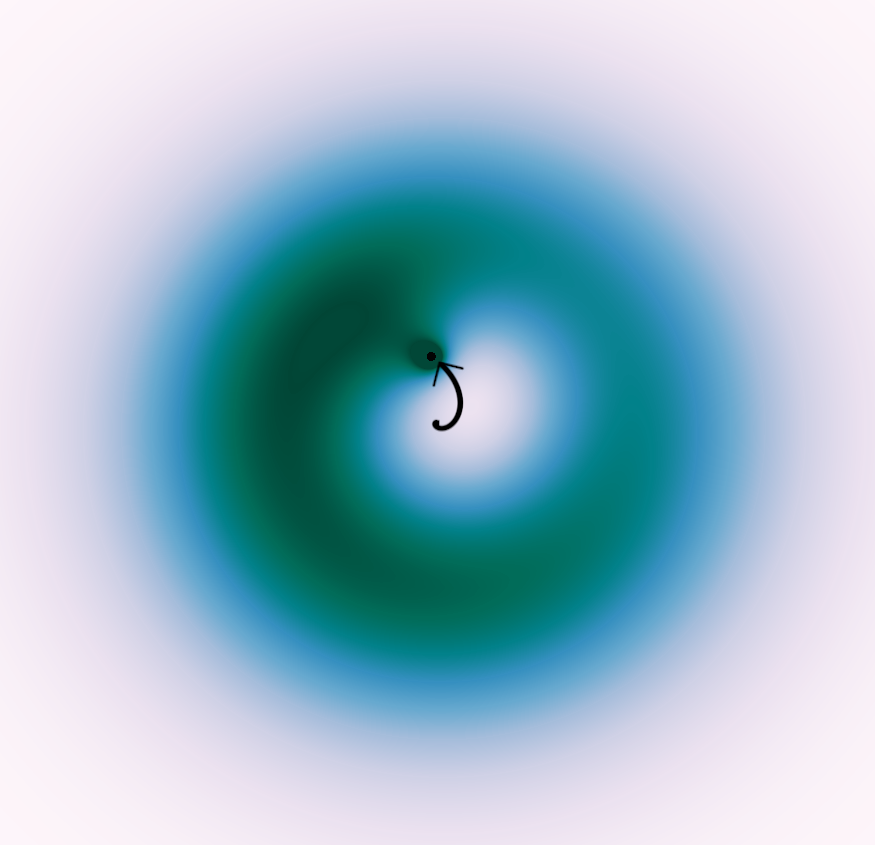}}
    \subfigure{\includegraphics[width=0.32\columnwidth]{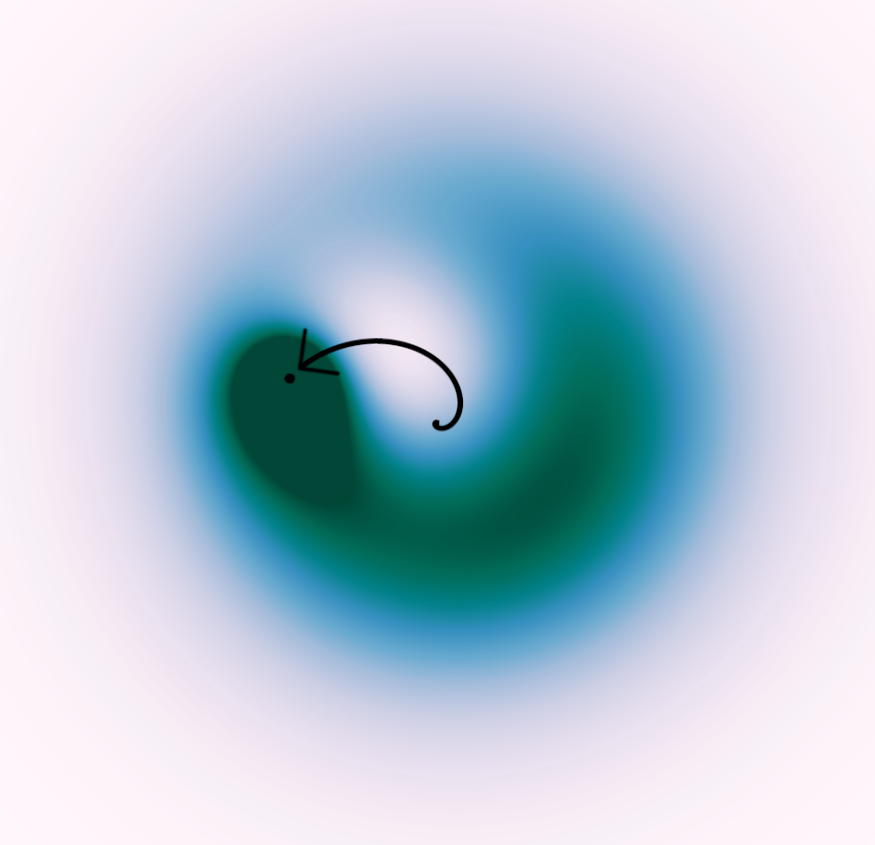}}
    \subfigure{\includegraphics[width=0.32\columnwidth]{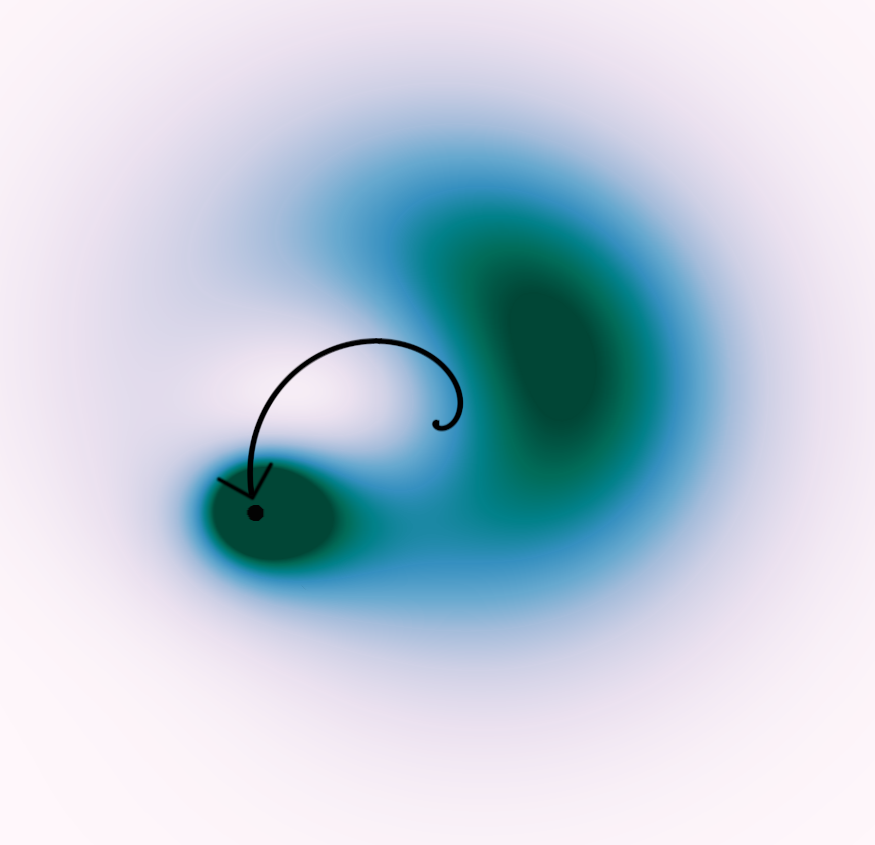}}
    \subfigure{\includegraphics[width=0.32\columnwidth]{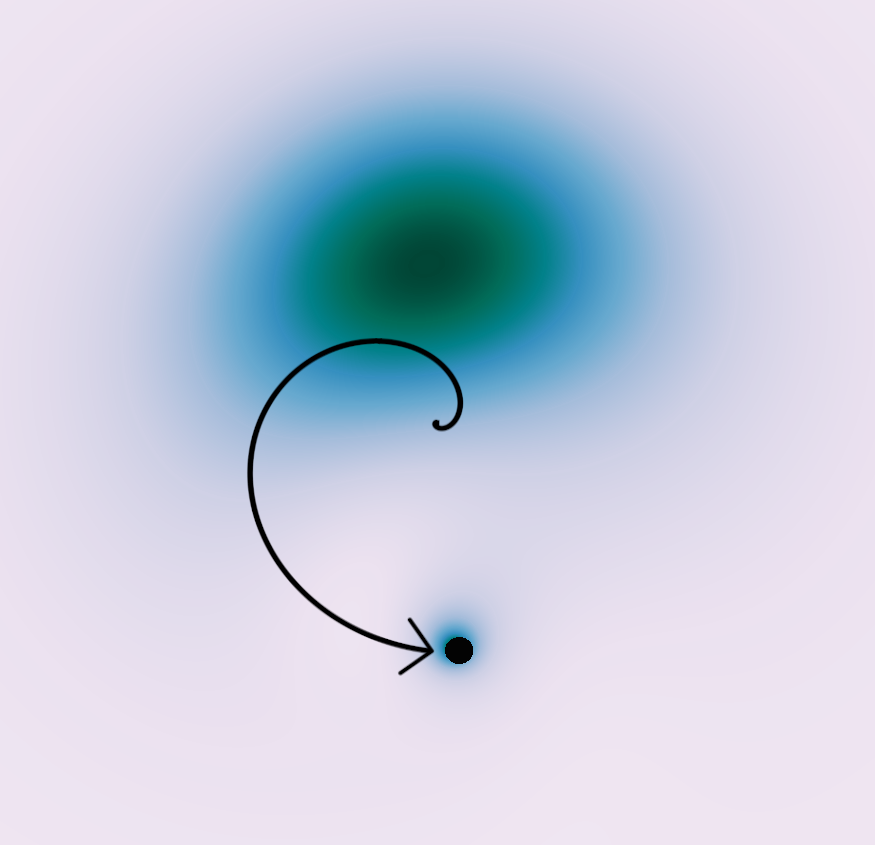}}
    \subfigure{\includegraphics[width=0.32\columnwidth]{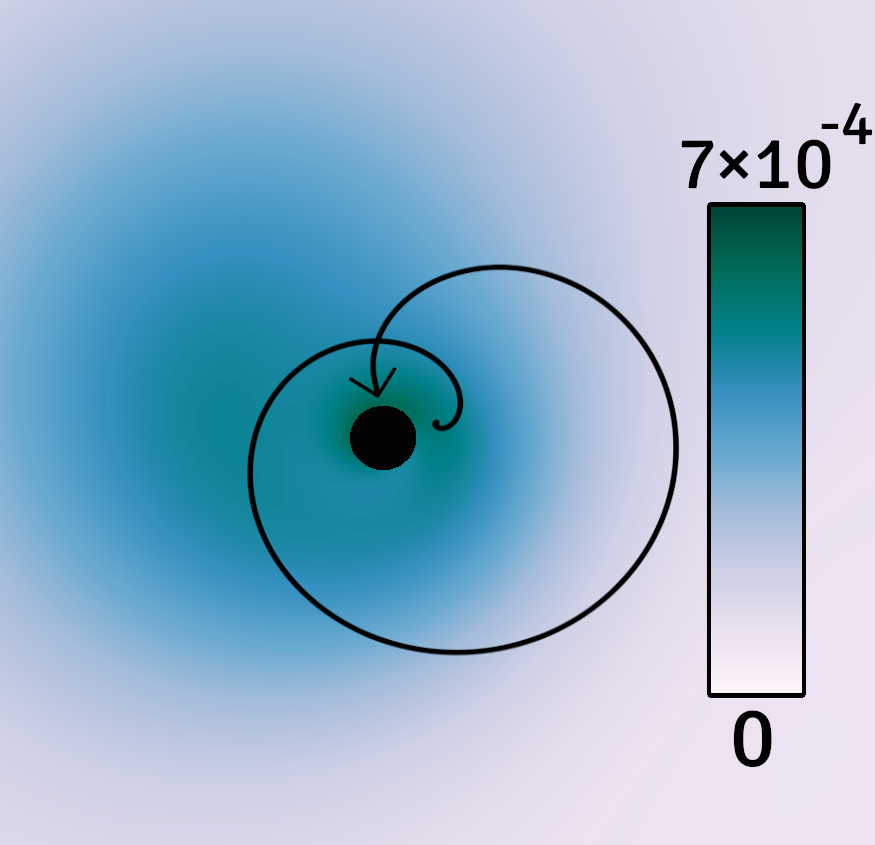}}
    \includegraphics[width=0.45\textwidth]{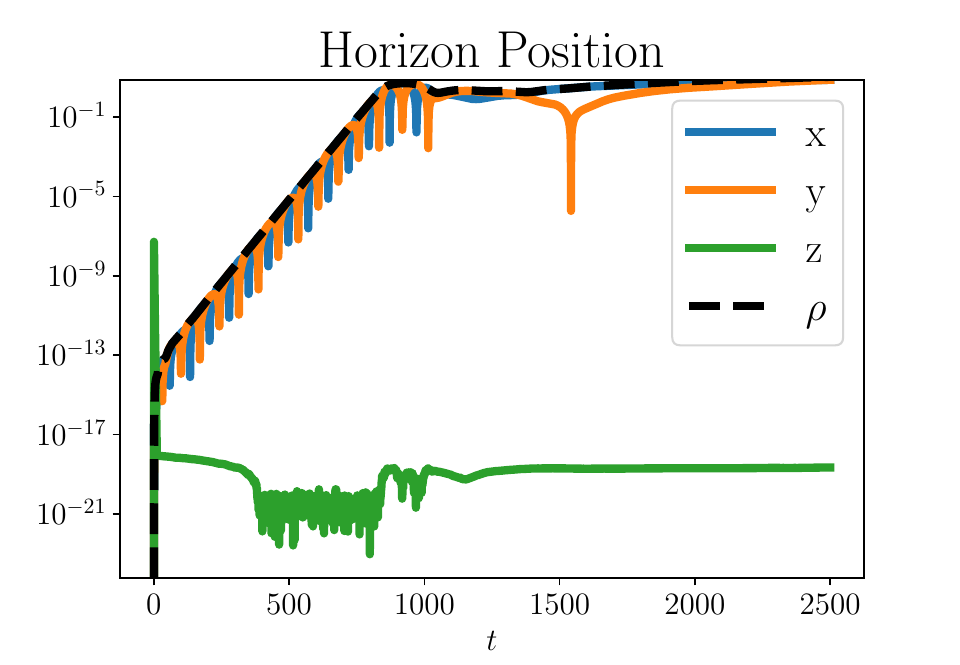}
    \caption{
    The BHSH configuration \textbf{C} spirals out from the origin, disrupts the scalar field and absorbs most of it when reaching high density regions.
    Top: snapshots, $t = 0, 748, 785, 805, 836 \text{ and } 1300$ -- left to right, top to bottom -- of the scalar field density in the $xy$ plane and
    the horizon trajectory (solid black arrows).
    Bottom: time evolution of the absolute value of the Cartesian coordinates and equatorial radius $\rho\equiv\sqrt{x^2+y^2}$ of the puncture, in log scale.
    }
    \label{fig2}
\end{figure}

Let us describe the evolution results focusing on the BHSH \textbf{C}, but similar results are obtained for configuration \textbf{B}. The main result is that from the start the BH starts to move around in an exponential outspiraling trajectory, in the direction of the angular momentum -- Fig.~\ref{fig2}.
When it reaches regions of higher scalar field density, it highly perturbs and disrupts the toroidal structure, with successive absorption phases, resulting in a large mass transfer from the scalar field to the BH~\footnote{For this simulation, the spiral has an approximately constant angular velocity of 0.043.}.
The scalar field density is given by $\rho_\Phi = n^\mu n^\nu T_{\mu\nu}$,
while its total energy is computed as a Komar integral~\cite{Gourgoulhon:2007ue,Gourgoulhon:2012ffd}:
$E_{\Phi}  = \int \left( 2T_{\mu\nu} - Tg_{\mu\nu} \right)(\partial_t)^\mu n^\nu \sqrt{\gamma} \ d^3x$,
where $n^\mu$ is the normal vector to the constant $t$ hypersurfaces,
$T_{\mu\nu}$ is the scalar energy-momentum tensor and $T$ its trace,
$g_{\mu\nu}$ is the 4-metric and $\gamma$ is the 3-metric's determinant.

\begin{figure}[h!]
\begin{center}
\includegraphics[width=0.45\textwidth]{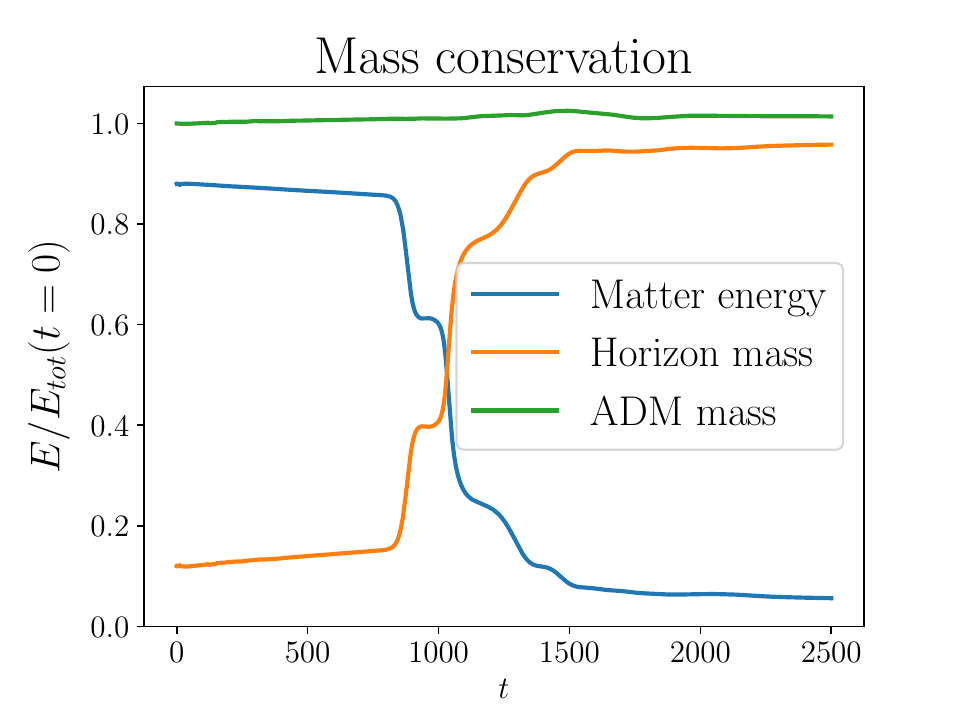}
\caption{\small Mass transfer between the BH and scalar field (configuration \textbf{C}).
The most dynamical phase occurs between $t \approx 700$ and $1500$.
}
\label{fig3}
\end{center}
\end{figure}

Figure~\ref{fig3} shows the energy exchange between the BH~\footnote{
We cannot rely on \texttt{QuasiLocalMeasures} to evaluate the BH mass, because it infers it from the Kerr relationship involving the BH area and angular momentum.
We thus compute the BH mass as the difference between the ADM mass and scalar field energy.
} and the scalar field.
The plot is rescaled so that the Arnowitt-Deser-Misner (ADM) mass~\footnote{
We numerically extract the ADM mass at 3 different radii, and extrapolate to infinity with $1/R$ linear dependence.
} is initially 1; it then remains relatively stable throughout the simulation, indicating that the system stays approximately isolated (gravitational waves and gravitational cooling appear negligible).  
The fraction of scalar energy to the total energy reduces from an initial $E_{\Phi}/E\approx 88\%$ to $5.5\%$ at $t = 2500$.

Qualitatively similar results are obtained for the angular momentum -- see Appendix A.
Quantitatively, the total angular momentum is not as well preserved as the energy throughout the evolution. This is known to be the case, even for the evolution of a single Kerr BH in a Cartesian grid~\cite{Baumgarte:2012xy,Mewes:2018szi,Mewes:2020vic}. 
The scalar angular momentum fraction drops from $\sim 99\%$ to $20\%$ corroborating the balding of the hairy BH. 
Additionally, the dimensionless spin of the horizon decreases  after the process to $\sim$0.63. The small scalar remnant supports that the final BH is approximately Kerr. It should again be transient, as it is unstable against superradiance triggered by $\Phi$, growing (smaller amounts of) hair. This suggests that this instability of very hairy BHsSH is a \textit{migration} to the almost bald region of their parameter space.

For configuration \textbf{A}, with the majority of its mass in the BH,  therefore not ``very hairy'', puncture movement is still observed  but without ever reaching physical values in the simulation timescales. The triggering of this motion is numerical, as it occurs even for a bald Kerr BH. For very hairy BHs, the simulations establish it becomes a physical outward motion, in agreement with the physics rationale presented above. This is not the case for almost bald BHs. But even if the difference is only quantitative, for sufficiently small hair, the timescales become long enough to compete with superradiant dynamics, so that this instability ceases to determine the evolution. Further details and examples of evolutions, both in and away from the very hairy regime are presented in Appendix B and elsewhere~\cite{Nicoules:2026}.

\smallskip
{\bf A cousin model -- BHs with resonant hair.}
Consider the same action as before, but now with matter Lagrangian
\begin{equation}
    \label{lag2}
    \mathcal{L}_{\rm m} =  -
        (D_\alpha \Psi)^* D^\alpha \Psi -  \mu^2 |\Psi|^2 (1 - 2 \lambda |\Psi|^2)^2  - 
        \frac{1}{4} F_{\alpha \beta} F^{\alpha \beta}
    \,,
\end{equation}
where $F_{\mu \nu} \equiv \nabla_\mu A_\nu - \nabla_\nu A_\mu$ is the Maxwell tensor, $A=A_\mu dx^\mu$ the 4-potential  to which the \textit{gauged} scalar field $\Psi$ couples (minimally) via $D_\mu \equiv \nabla_\mu  + i q A_\mu$, 
and $q$ is the gauge coupling constant. Again, we take $\mu=1$. A solution of this model is the Einstein-Maxwell Reissner-Nordstr\"om (RN) BH and $\Psi=0$. This BH is prone to superradiant \textit{scattering} by $\Psi$ when scalar field modes of frequency $\omega$ obey $\omega<q U(r_H)$, with $U(r_H)$ the horizon electrostatic potential. Additionally, under the \textit{resonance} condition $\omega=q U(r_H)$, model~\eqref{lag2} allows \textit{BHs with resonant (scalar) hair} (BHsRH)~\cite{Herdeiro:2020xmb,Hong:2020miv}. 

\textit{Spherical} and static BHsRH were found numerically in \cite{Herdeiro:2020xmb,Hong:2020miv}. Here, we take an \textit{Ansatz} in isotropic coordinates
$ds^2 = -e^{2\mathcal{F}_0} S_0^2{S_1^{-2}} dt^2 + e^{2\mathcal{F}_1} S_1^4 [dr^2+r^2 (d\theta^2+\sin^2\theta d\varphi^2)]$
where  $\mathcal{F}_0$ and $\mathcal{F}_1$ are radial functions and $    S_0 \equiv 1 - {r_H}/{r}$, $  S_1 \equiv 1 + r_H/{r}$. The matter \textit{Ans\"atze} are $\Psi = \psi(r) e^{-i \omega t}$ and  $A = U(r) dt$.   Then, as for BHsSH, the data computed in the exterior region can be extended to the whole radial range. 

The parameter space of this model has been (partially) discussed in the literature~\cite{Herdeiro:2020xmb,Hong:2020miv}. As for BHsSH, for vanishing BH size (implying $r_H\rightarrow 0$), one finds (gauged) boson stars~\cite{Jetzer:1989av,Pugliese:2013gsa}. In their vicinity very hairy BHs exist. But unlike BHsSH, there is no $\Psi\rightarrow 0$ limit; RN BHs are disconnected from BHsRH. This can be understood from a linear $\Psi$ analysis around RN BHs: no superradiant \textit{instabilities} or bound states exist, under the resonance condition~\cite{Hod:2012wmy}. But non-linear bound states, \emph{$Q$-clouds}, on an RN background exist~\cite{Herdeiro:2020xmb}, amounting to studying a decoupling limit of~\eqref{lag2}; this justifies the self-interaction term therein, with coupling $\lambda$. The model thus admits very hairy, but not almost bald, BHs.

\begin{figure}[tbh!]
\begin{center}
\includegraphics[width=0.45\textwidth]{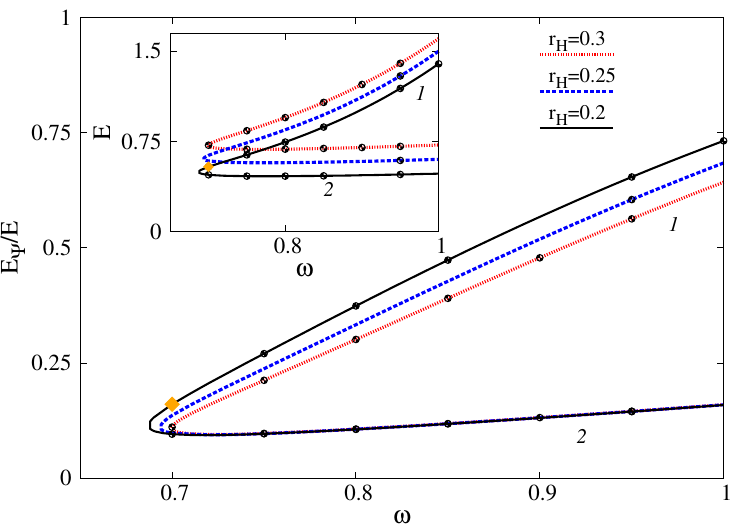}
\caption{\small Fraction of scalar energy (main panel) and total spacetime energy $E$ (inset) vs~$\omega$ for sequences of solutions with different $r_H$. Two connected branches arise, i.e.~two solutions for the same $\omega$. The black dots and orange diamond mark the solutions numerically evolved.}
\label{fig4}
\end{center}
\end{figure}

Fixing illustrative values,  $\lambda=2500,q=12$, Fig.~\ref{fig4}
shows the total energy $E$ (inset), as well as the fraction of the energy in the scalar field $E_\Psi/E$ for sequences of solutions with fixed $r_H$, in terms of $\omega$~\footnote{We remark that there is also electromagnetic energy outside the horizon.}. The very hairy solutions always occur in the top branch (labeled 1) of the solutions curve. In~\cite{Herdeiro:2020xmb} the quantity $h\equiv 1-Q_H/Q$ was proposed as measure of hairiness in this model, where $Q_H$ and $Q$ are the horizon and total charges, respectively. This quantity becomes close to unity along branch 1, as $\omega\rightarrow 1$. 

\begin{figure}[t]
    \centering
    \subfigure[\ Initial BH with resonant hair]{\includegraphics[width=0.45\columnwidth]{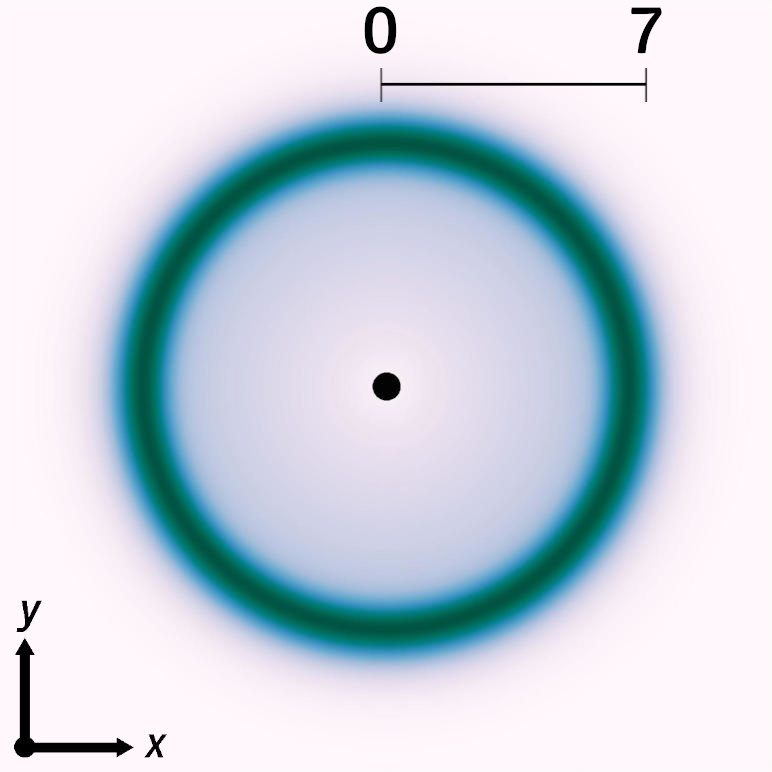}} \hfill
    \subfigure[\ Horizon moves away]{\includegraphics[width=0.45\columnwidth]{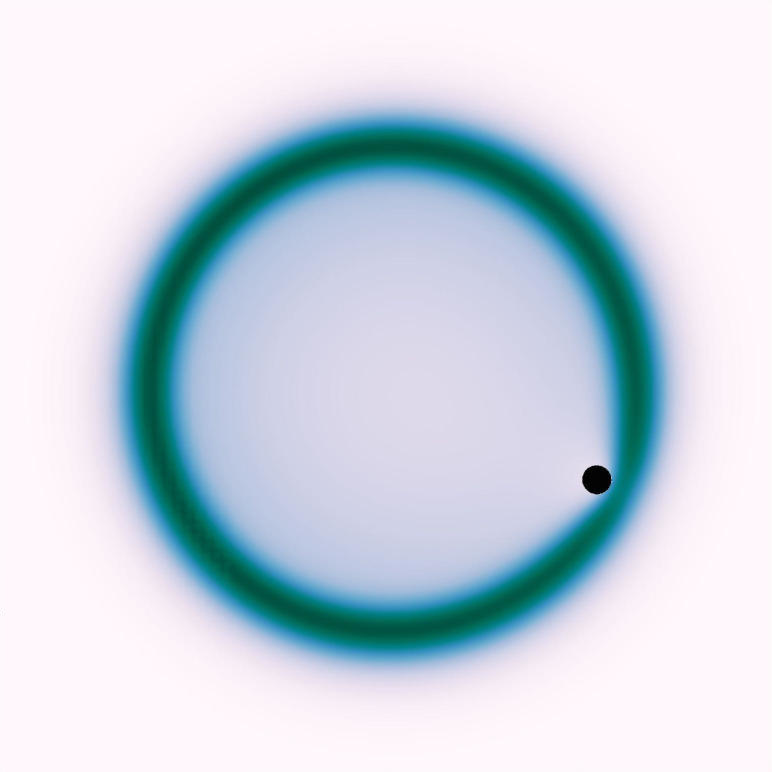}} \\
    \subfigure[\ Horizon and boson star split]{\includegraphics[width=0.45\columnwidth]{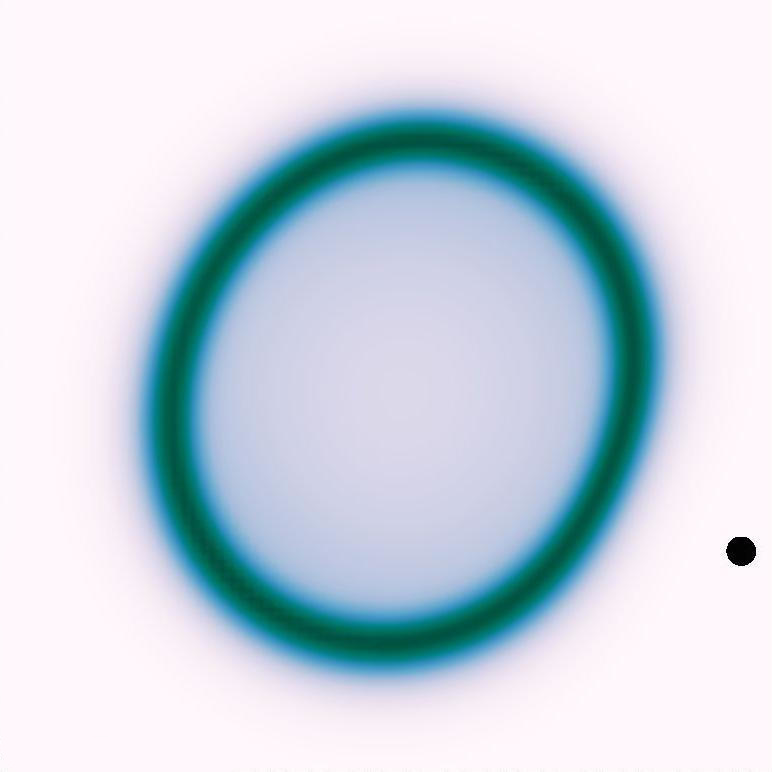}} \hfill
    \subfigure[\ Oscillating boson star remnant]{\includegraphics[width=0.45\columnwidth]{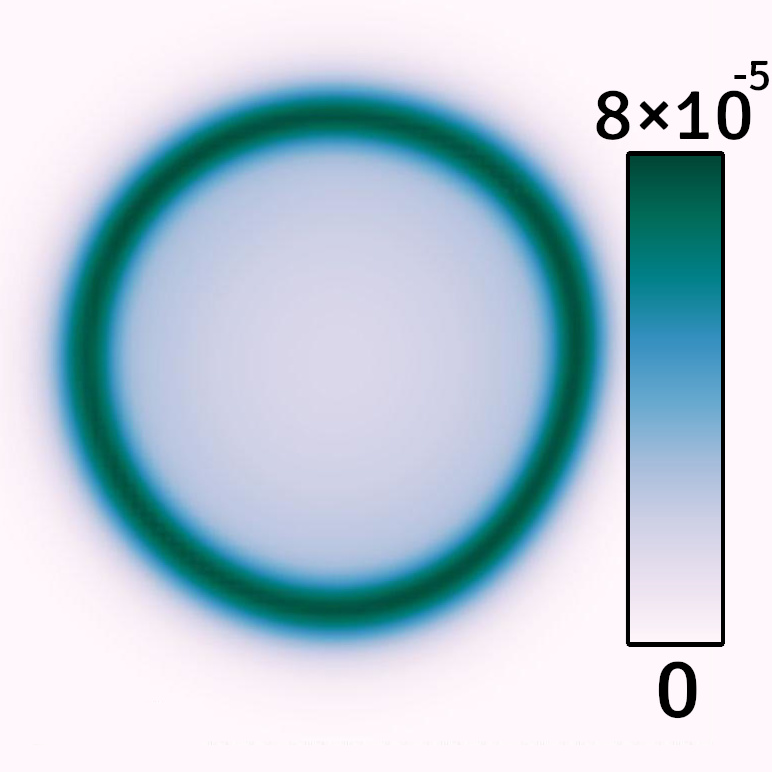}}
    \caption{Snapshots (for $t =  0, 1720, 1760 \text{ and } 1784$) of the density of the scalar field in the $xy$ plane. The BH is expelled by the hair, leaving behind a stable boson star with non-zero linear momentum. The solution evolved corresponds to the orange diamond in Fig.~\ref{fig4}.}
    \label{fig5}
\end{figure}

The maximum of the scalar density is attained at some distance from the horizon -- Fig.~\ref{fig5} (first panel). A very hairy system can therefore be understood as a small charged particle inside a large (spherical) charged boson star. In a Newtonian analysis of such a system, considering an electrically charged star with mass and charge density $\rho_M$ and $\rho_Q$ near the center, a particle with mass $m$ and charge $q$ is in an unstable equilibrium at the central point if $q\rho_Q>m\rho_M$. The following evolutions support that a relativist analog of this condition applies for very hairy BHsRH.

{\bf Non-linear dynamics of BHsRH.}
RN BHs are linearly stable against scalar perturbations in the model described by Eq.~(\ref{lag2}), but numerical simulations have provided evidence that they turn out to be non-linearly (in the scalar field) unstable, with the end point being BHsRH~\cite{Zhang:2023qtn}. Such simulations suggest the stability of BHsRH under \textit{spherical} dynamics, which was assumed. This raises the question whether they are stable under more generic dynamics. 
To address this question, we again make use of the \texttt{Einstein Toolkit} infrastructure for performing evolutions as detailed above, including also the thorn \texttt{MagnetoScalar}~\cite{Jaramillo2024,Zilhao:2015tya}. Further details can be found in a follow-up paper~\cite{Ferreira:2026}.
Details about the initial data, evolution equations and convergence studies, of this model and the previous one, are presented in Appendices C-E.

The time evolutions of very hairy BHsRH, corresponding to the top branch in Fig.~\ref{fig4}, show a clear pattern. After some time, the horizon starts moving away from the central equilibrium point, splitting from the surrounding scalar environment and eventually being ejected from it. Unlike the case of BHsSH, in which the toroidal scalar environment was mostly absorbed and no solitonic isolated object was seen to survive, in this case an oscillating boson star is observed as a remnant -- Fig.~\ref{fig5}. The remnant star has linear momentum in the opposite direction to the ejected horizon. This dynamics is nonspherical and outside the scope of the simulations in~\cite{Zhang:2023qtn}. One reason for the different fate in the two models may be the dynamical instability of the toroidal spinning boson stars of model~\eqref{action}~\cite{Sanchis-Gual:2019ljs} as opposed to the dynamical robustness of gauged spherical boson stars of model~\eqref{lag2}~\cite{Jaramillo:2024shi}. 

We have also considered the evolutions of less hairy solutions, as the ones in the lower branches in Fig.~\ref{fig4}, and with higher frequency. These were also found to be unstable, but  instead of a splitting of these gravitational atoms, the scalar environment collapses into the horizon, being absorbed by it. More details about this case will be presented in~\cite{Ferreira:2026}. 
Unlike the case of BHsSH we found no stable/long-lived solutions, which may be related to the absence of almost bald BHsRH. 

\smallskip
{\bf Remarks and generality.}
The synchronization condition allows a spectrum of scalar structures around horizons with different quantum numbers~\cite{Herdeiro:2015gia,Wang:2018xhw,Delgado:2019prc,Kunz:2019bhm}. In the test field approximation, these scalar structures have a hydrogenlike spectrum justifying the terminology ``gravitational atoms''. This has been seen for both long-lived quasi-bound states (typically connected to superradiance~\cite{Baumann:2019eav}), and for stationary bound states~\cite{Hod:2012px}. Away from the test field approximation, the stationary states become hairy BHs, and in the very hairy region, a small horizon inside a massive boson star. We have shown the latter tend to split, with a model dependent fate, tentatively related to the stability of the bosonic environment. 

One may inquire if this instability is generic within the larger family of BHsSH/BHsRH, including excited states (say with larger $m$)~\footnote{The two models can be related by considering a higher dimensional setting~\cite{Brihaye:2023vox}.}. This remains to be seen.
One may anticipate that appropriate self-interactions have an important impact as they mitigate or fully quench the nonaxisymmetric instability of spinning boson stars, making them less toroidal~\cite{DiGiovanni:2020ror,Siemonsen:2020hcg}. However, from a geodesic approximation, the origin is never a stable equilibrium point~\cite{Delgado:2021jxd}, suggesting the splitting remains~\footnote{For the case of BHsRH, a similar instability was briefly commented in~\cite{Sanchis-Gual:2025uvm}.}.

A potentially different case is that of BHs with synchronized Proca hair~\cite{Herdeiro:2016tmi}. Proca stars are spheroidal, rather than toroidal, and the nonaxisymmetric instability has not been seen here~\cite{Sanchis-Gual:2019ljs}. Additionally, from a geodesic approximation the origin can be a stable equilibrium point, at least for co-rotating motions~\cite{Delgado:2021jxd}. Thus, the dynamical fate of very hairy Proca BHs deserves a detailed analysis, which is underway.

\bigskip

{\bf Acknowledgements.}
This work is supported by the Center for Research and Development in Mathematics and Applications (CIDMA) (\url{https://ror.org/05pm2mw36}) under the Portuguese Foundation for Science and Technology 
(FCT -- Fundaç\~ao para a Ci\^encia e a Tecnologia, \url{https://ror.org/00snfqn58}), Grants UID/04106/2025 (\url{https://doi.org/10.54499/UID/04106/2025}) and UID/PRR/04106/2025 (\url{https://doi.org/10.54499/UID/PRR/04106/2025}), as well as the projects: Horizon Europe staff exchange (SE) programme HORIZON-MSCA2021-SE-01 Grant No.\ NewFunFiCO-101086251;  2022.04560.PTDC (\url{https://doi.org/10.54499/2022.04560.PTDC}) and 2024.05617.CERN (\url{https://doi.org/10.54499/2024.05617.CERN}).
M.Z.\ is funded by FCT through project 2022.00721.CEECIND 
(\url{https://doi.org/10.54499/2022.00721.CEECIND/CP1720/CT0001}). 
J.F.\ is funded by FCT through project 2023.04333.BD (\url{https://doi.org/10.54499/2023.04333.BD}).
The authors thankfully acknowledge computational resources from RES provided by BSC (MareNostrum) through projects FI-2024-2-0012, FI-2024-3-0007, POR021PROD, and by IFCA (Altamira) through project FI-2025-1-0011.
Computational resources were also provided via FCT through project 2024.07872.CPCA.A2 at Deucalion supercomputer, jointly funded by EuroHPC JU and Portugal, and at MareNostrum through project 2024.07059.CPCA.A3 (DOI: 10.54499/2024.07059.CPCA.A3 \url{https://doi.org/10.54499/2024.07059.CPCA.A3}).
This work was granted access to the HPC resources of MesoPSL financed by the Region Ile de France and the project Equip@Meso (reference ANR-10-EQPX-29-01) of the programme Investissements d'Avenir supervised by the Agence Nationale pour la Recherche.

\smallskip
{\bf Data availability.}
The data that support the findings of this article are not publicly available upon publication because it is not technically feasible and/or the cost of preparing, depositing, and hosting the data would be prohibitive within the terms of this research project. The data are available from the authors upon reasonable request.


\bibliography{letterbhlr}


\appendix

\section{Appendix A - Angular momentum dynamics for Black Holes with Synchronized Hair}
\label{apA}
\begin{figure}[h!]
\begin{center}
\includegraphics[width=0.45\textwidth]{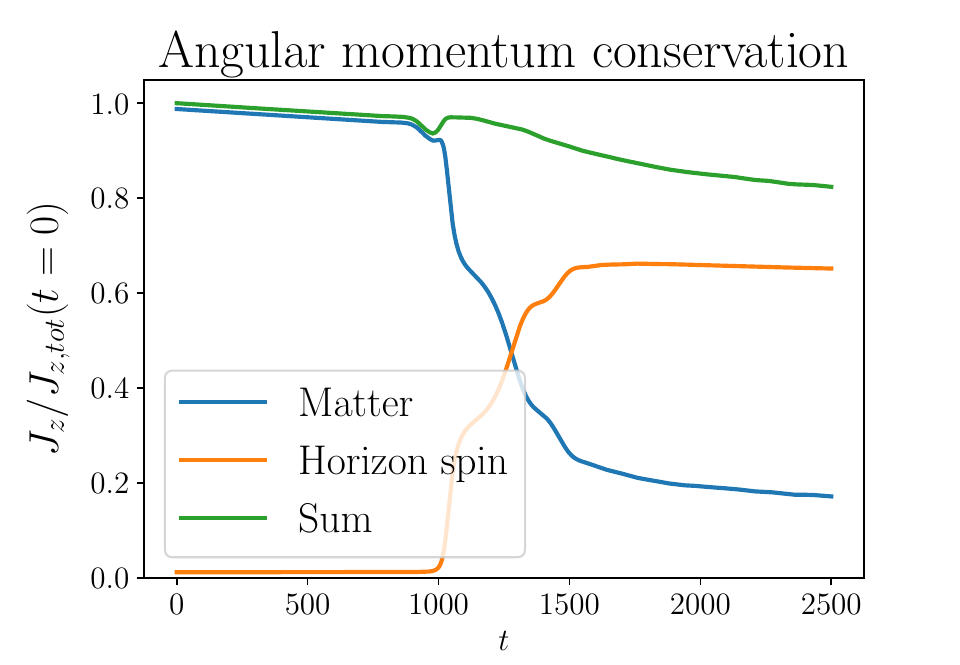}
\caption{\small Angular momentum transfer between the BH and scalar field for BHSH configuration \textbf{C}, with time in units of $\mu$. }
\label{figa1}
\end{center}
\end{figure}

%
Fig.~\ref{figa1} shows the exchange of angular momentum ($z$-component) between the scalar field and the black hole (BH), for the BH with synchronized hair (BHSH) configuration \textbf{C}.
As for the mass, for the scalar field this is computed as a Komar integral with respect to the relevant vector field, $\partial_\varphi$,
\begin{align}
    J_{z,\Phi}    &= -\int \left( T_{\mu\nu} - \dfrac{1}{2}Tg_{\mu\nu} \right)(\partial_\varphi)^\mu n^\nu \sqrt{\gamma} \ d^3x \\
                    &= \int \left( xp_y - yp_x \right)\sqrt{\gamma} \ d^3x \ ,
\end{align}
where $p_i = -T_{\mu\nu}\gamma^{\mu}_{~~i}n^{\nu}$ is the momentum density and $\gamma_{ij}$ is the 3-metric.
Here, we have direct access to the BH angular momentum, so the plot is rescaled by the initial sum of the scalar field and BH contributions.
For angular momentum, the initial value is about 0.91, and the sum decreases a bit more significantly, ending at 82\%.
The scalar contribution to the angular momentum proportion drops from $\approx 99\%$ to $20\%$.
Analyzing gravitational waves seems to indicate that loss of energy and angular momentum through gravitational wave emission is not significant.

\begin{figure}[h!]
\begin{center}
\includegraphics[width=0.45\textwidth]{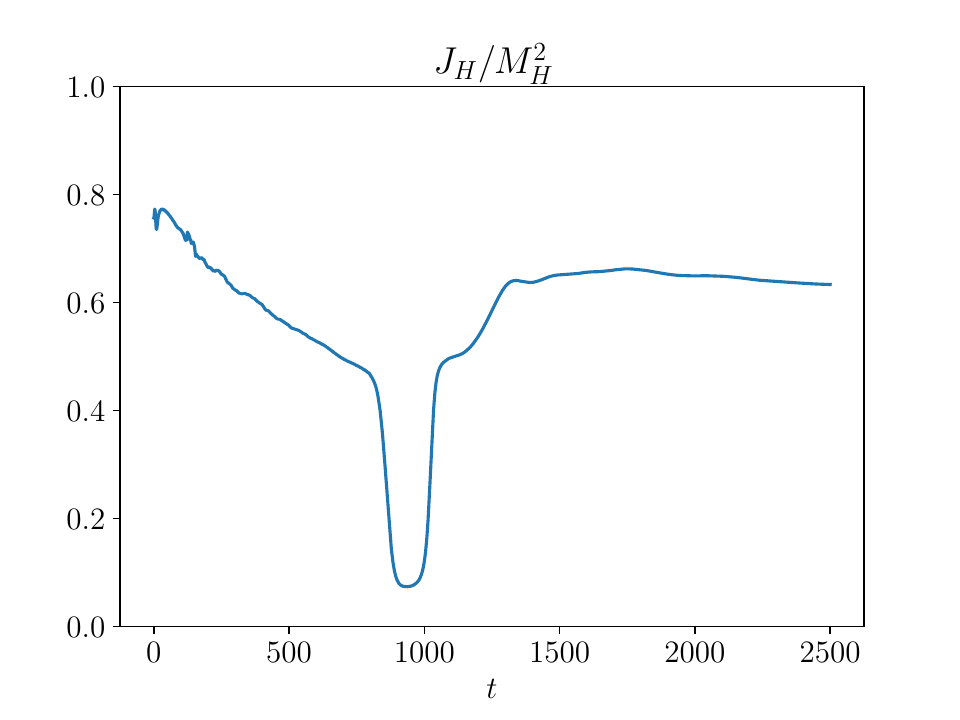}
\caption{\small Evolution of the BH dimensionless spin for BHSH configuration \textbf{C}, with time in units of $\mu$.}
\label{figa2}
\end{center}
\end{figure}

The difference in mass and angular momentum transfer dynamics translates into a non-trivial evolution of the dimensionless spin of the horizon -- Fig.~\ref{figa2} -- but the BH ends up spinning more slowly after the process.

In our simulation, although the vector fields associated with stationarity and axisymmetry are not Killing fields, we still use them as proxies.
Our results show that this approximation is still helpful and insightful, although it may contribute to the lack of conservation and can explain the turmoil around $t\mu=1000$.

\section{Appendix B - Evolutions for cases  A and B of BHs with synchronized hair}
\label{apB}

\begin{figure}[tbph]
    \centering
    \subfigure[\ Configuration \textbf{A}]{\includegraphics[width=0.45\textwidth]{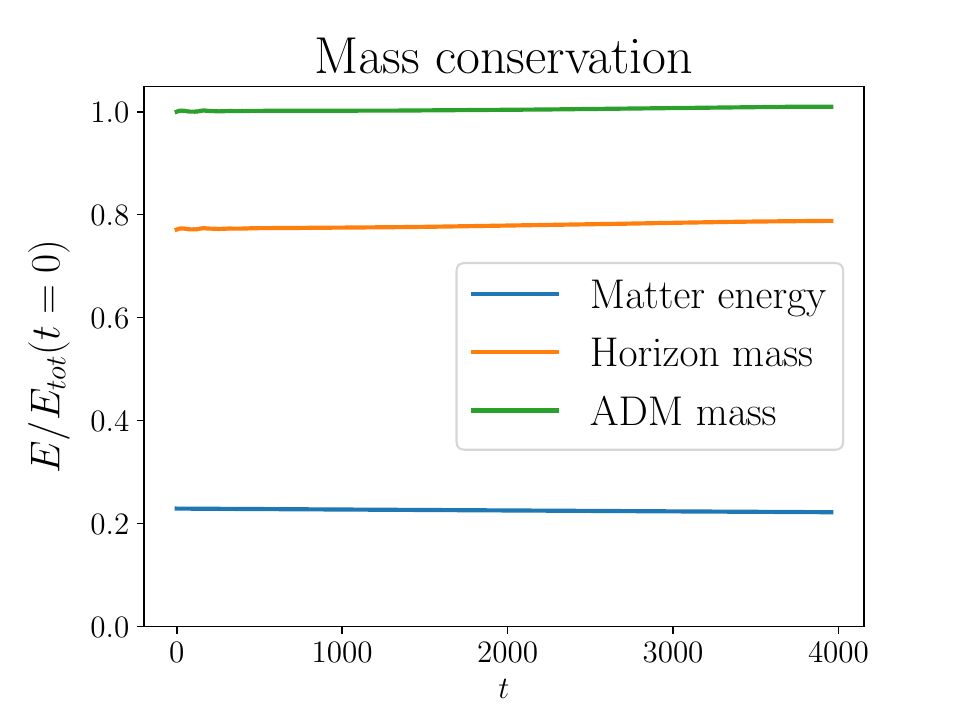}}
    \subfigure[\ Configuration \textbf{B}]{\includegraphics[width=0.45\textwidth]{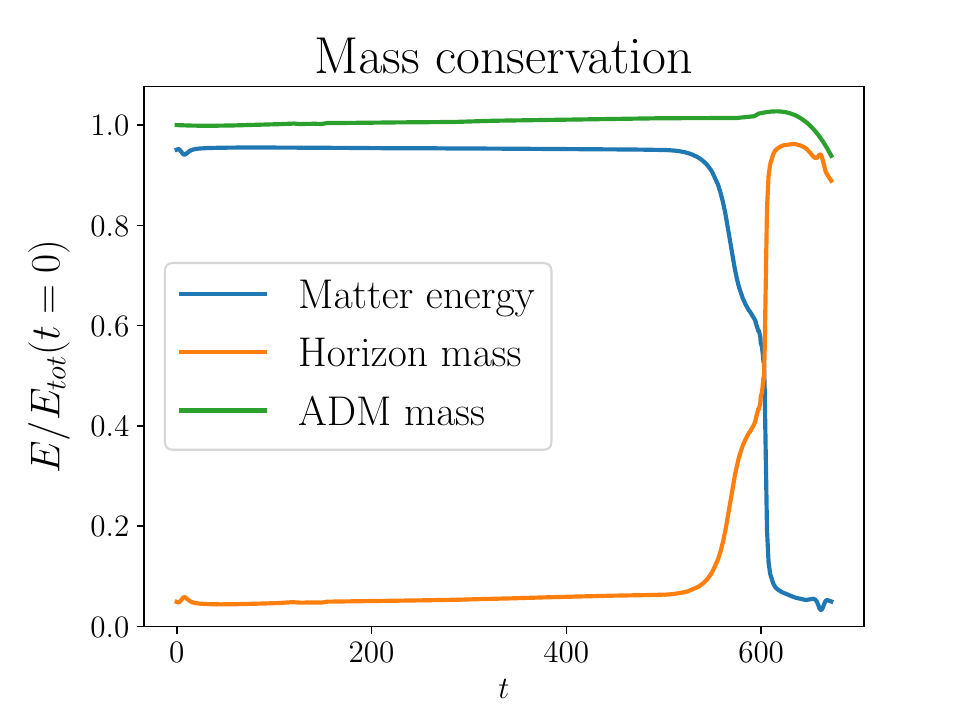}}
    \caption{Mass transfer between the BH and the scalar field for BHsSH configurations \textbf{A} and \textbf{B}, with time in units of $\mu$.
    The not ``very hairy'' configuration \textbf{A} shows no instability on long time scales.
    On the other hand, configuration \textbf{B} shows clear signs of the same type of instability as configuration \textbf{C}.
    }
    \label{figb1}
\end{figure}

We present here results for BHsSH configurations \textbf{A} and \textbf{B}.
We illustrate the respective stable and unstable behaviors in Fig.~\ref{figb1} by showing the mass transfer between the BH and the scalar field.
Note that configuration \textbf{B} is more difficult to tackle numerically, especially at late times.
Even though refined numerical treatments may be implemented in the future, the current results already support our point regarding the instability of ``very hairy'' configurations.

We would like to add that our results were corroborated by the work~\cite{Carretero:2025eqh} that appeared in the meantime (analyzing one of the models in our letter). Moreover, it supports the idea that BHs with less hair do not exhibit the same instability.

\section{Appendix C - Initial Data}
\label{apC}
\begin{figure*}[p]
    \centering
    \subfigure[\ $F_0$]{\includegraphics[width=0.35\textwidth]{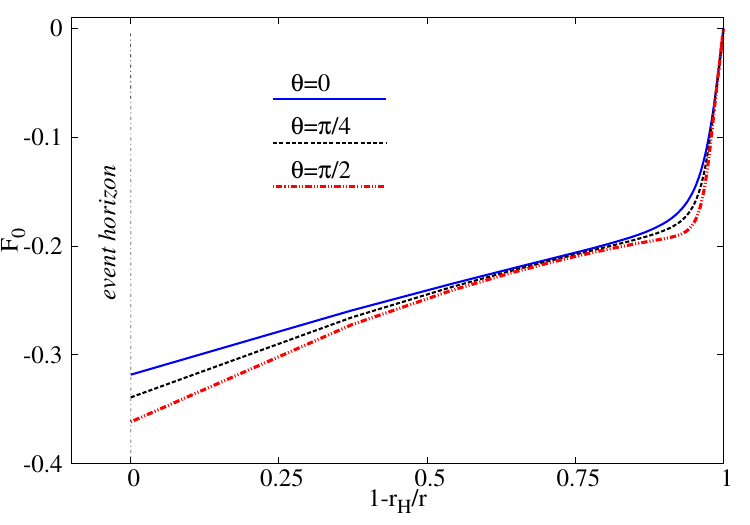}}
    \subfigure[\ $F_1$]{\includegraphics[width=0.35\textwidth]{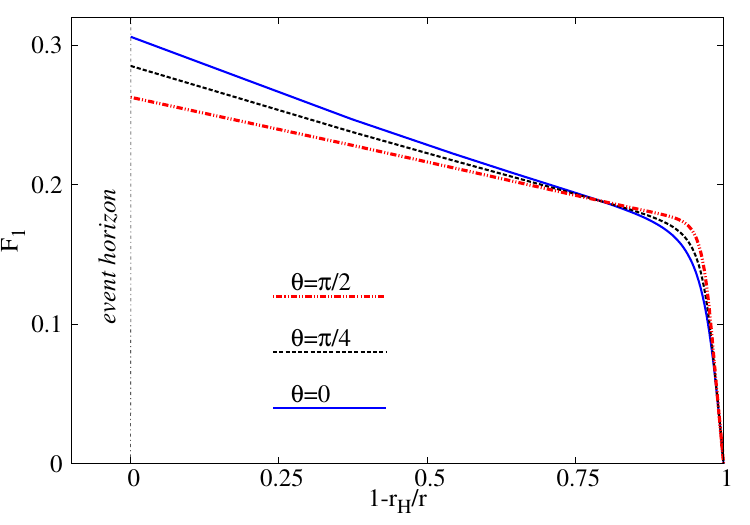}}
     \subfigure[\ $F_2$]{\includegraphics[width=0.35\textwidth]{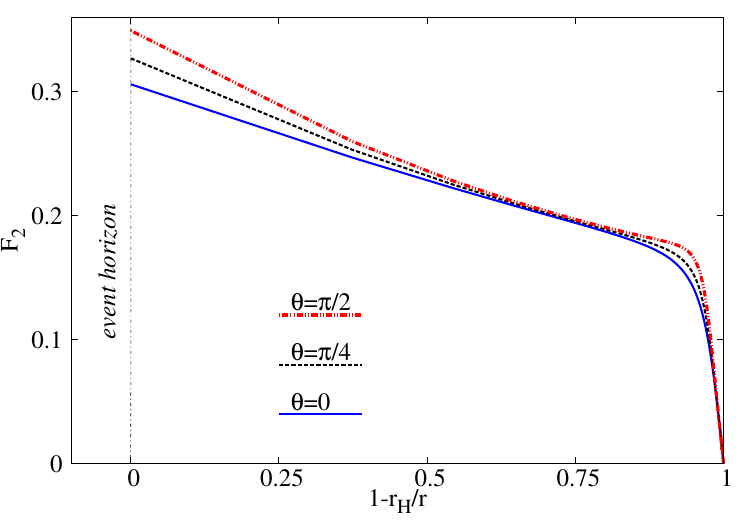}}
    \subfigure[\ $W$]{\includegraphics[width=0.35\textwidth]{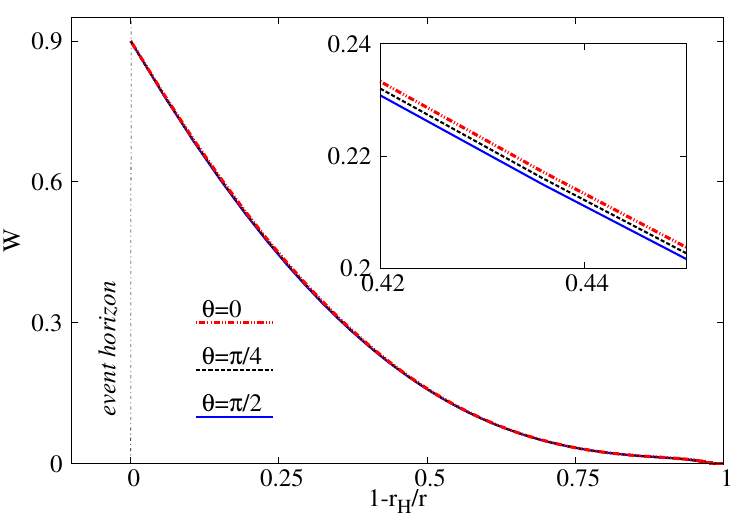}}
     \subfigure[\ $\phi$]{\includegraphics[width=0.35\textwidth]{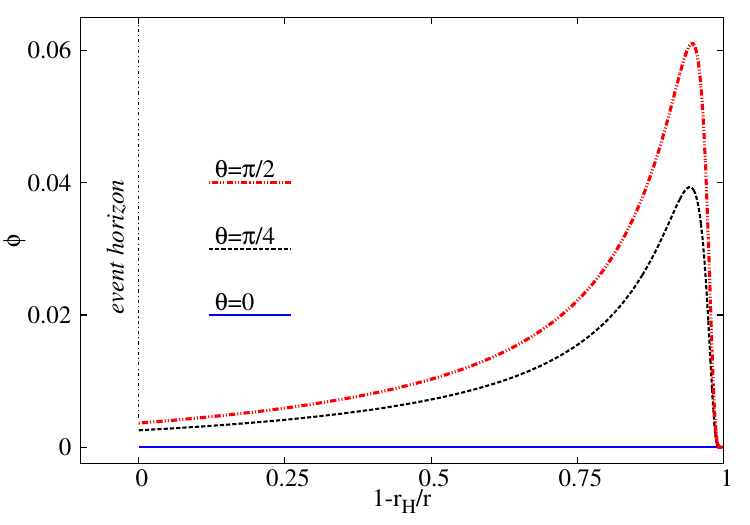}}
    \caption{Profile functions for the BH with synchronized hair solution evolved in Fig.~\ref{fig2}. 
    }
    \label{figc1}
\end{figure*}

In Fig.~\ref{figc1} we show the profiles of the several metric and scalar field functions for the BHSH evolved in Fig.~\ref{fig2} (configuration {\bf C} in the table in the main text), using the notation therein.

\begin{figure*}[p]
    \centering
   \includegraphics[width=0.35\textwidth]{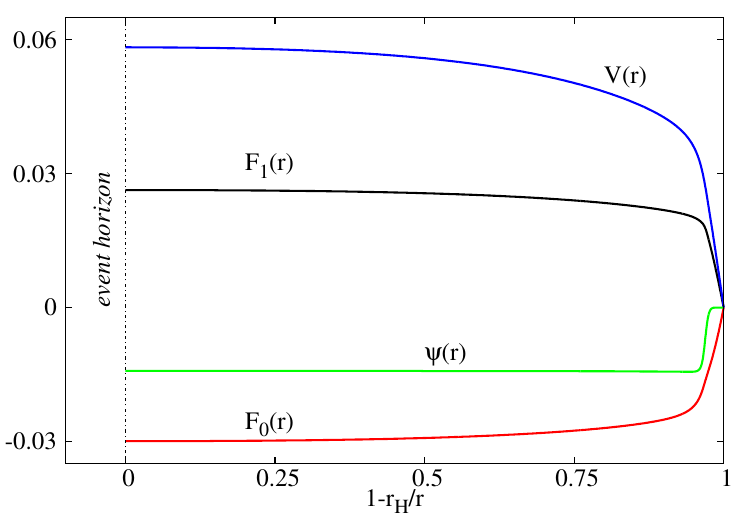}
    \caption{Profile functions for the BH with resonant hair solution evolved in Fig.~\ref{fig5}. 
    }
    \label{figc2}
\end{figure*}

In Fig.~\ref{figc2} we show the profiles of the several metric, scalar and gauge field functions for the BH with resonant hair (BHRH) evolved in Fig.~\ref{fig5}, using the notation therein. This solution is highlighted with an orange diamond in Fig.~\ref{fig4}, and has $r_H=0.2,\omega=0.7$ (in units of $\mu$), in the upper branch.

\section{Appendix D - Equations of Motion}
\label{apD}

To numerically integrate the preceding equations, we employ the $3+1$ decomposition of the field equations. The spacetime metric is expressed as
\begin{equation}
    \label{eq:3+1-line-element}
    ds^2 = -\alpha^2 dt^2 + \gamma_{ij}(dx^i + \beta^i dt)(dx^j + \beta^j dt) \,,
\end{equation}
where latin indices run on spatial coordinates, $\alpha$ is the lapse function, $\beta^i$ is the shift vector, and $\gamma_{\mu \nu}$ is the induced metric on the spatial hypersurfaces, defined by
\begin{equation}
    \label{eq:3-metric}
    \gamma_{\mu \nu} \equiv g_{\mu \nu} + n_\mu n_\nu \,,
\end{equation}
with $n_\mu = (-\alpha,0,0,0)$ representing the future-directed unit normal vector.
We write ${}^{(3)}D_i$ the covariant derivative compatible with $\gamma_{ij}$, defined in general as
\begin{align}
    \label{eq:3D-derivative}
    \begin{split}
        & {}^{(3)}D_i \mathcal T^{\mu_1 \ldots \mu_m}{}_{\nu_1 \ldots \nu_n} \equiv \\
        & \gamma^{\mu_1}{}_{\alpha_1} \ldots \gamma^{\mu_m}{}_{\alpha_m}
        \gamma^{\beta_1}{}_{\nu_1}  \ldots \gamma^{\beta_n}{}_{\nu_n}
        \gamma^\rho{}_i
        \nabla_\rho \mathcal T^{\alpha_1 \ldots \alpha_m}{}_{\beta_1 \ldots \beta_n} \,.
    \end{split}
\end{align}

Finally, the extrinsic curvature is given by 
\begin{equation}
    \label{eq:extr_curvature}
    K_{ij} \equiv - \dfrac{1}{2\alpha} \left( \partial_t - \beta^k \partial_k \right) \gamma_{ij} \,,
\end{equation}
and we note its trace $K \equiv \gamma^{ij}K_{ij}$.
Since the evolution equations resulting from the 3+1 projection of Einstein's equations are second order in time, $K_{ij}$ can be seen as a reduction variable.
It is also convenient to perform a first-order reduction in time for the scalar field, which results in the introduction of its conjugate momentum $K_\Phi$, such that (replace $\Phi$ by $\Psi$ for the BHsRH)
\begin{equation}
    \label{eq:K_phi}
    K_\Phi \equiv - \dfrac{1}{2\alpha} \left( \partial_t - \beta^k \partial_k \right) \Phi  \,.
\end{equation}

It is relevant to note here that we do not evolve directly the ansatz functions presented in the main text.
Rather, we compute the 3+1 quantities which were just presented, in Cartesian coordinates $x, y, z$ and with the associated coordinate tensor basis, such that 
the corresponding radius is the (quasi-) isotropic radius introduced in the main text.
%
The metric sector is evolved following the well-established BSSN (Baumgarte-Shapiro-Shibata-Nakamura) formalism~\cite{PhysRevD.59.024007, PhysRevD.52.5428}, which is a 3+1-based formulation of Einstein's equations well-suited for numerical evolutions, but relies on a slightly different set of evolution variables.
The conversion from standard 3+1 fields to BSSN variables is handled directly within the evolution \textit{thorn} from the \texttt{Einstein Toolkit}.
An explicit form of the evolution system in the presence of matter is given in Appendix A of~\cite{Cunha:2017wao}.
For the gauge evolution, we rely on standard $1+\log$ slicing and $\Gamma$-driver shift (in its first-order form~\cite{vanMeter:2006vi}):
\begin{equation}
    \partial_t \alpha = \beta^k\partial_k\alpha - 2\alpha K \,,
\end{equation}
\begin{equation}
    \partial_t \beta^i = \beta^k\partial_k\beta^i +  \dfrac{3}{4}\tilde{\Gamma}^i - \eta \beta^i \,,
\end{equation}
where $\tilde{\Gamma}^i$ is the conformal connection vector from the BSSN system and $\eta$ is a damping parameter.
 
\subsection{D1. BHs with Synchronized Hair}

The Lagrangian given in Eq. 1 of the main text, representing a complex massive scalar field minimally coupled to gravity, yields the corresponding energy-momentum tensor to be used in Einstein's equations:
\begin{equation}
    T_{\mu\nu} = \nabla_{\mu}\Phi^*\nabla_{\nu}\Phi + \nabla_{\nu}\Phi^*\nabla_{\mu}\Phi - g_{\mu\nu} \left( \nabla_\alpha\Phi^* \nabla^\alpha\Phi + \mu^2 \Phi^* \Phi \right) \,.
\end{equation}

After performing the 3+1 decomposition of the Klein-Gordon equation, the following evolution equations are obtained~\cite{Cunha:2017wao}:
\begin{equation}
    \label{eq:evo_Phi}
    \partial_t \Phi = \beta^i\partial_i\Phi - 2\alpha K_\Phi \,,
\end{equation}
\begin{align}
    \begin{split}
    & \partial_tK_\Phi = \beta^i\partial_iK_\Phi  \\
    & + \alpha \left( KK_\Phi - \dfrac{1}{2}\gamma^{ij} ~{}^{(3)}D_i\partial_j \Phi + \dfrac{1}{2}\mu^2\Phi \right)
    - \dfrac{1}{2} \gamma^{ij}\partial_i\alpha\partial_j\Phi \,.
    \end{split}
\end{align}

\subsection{D2. BHs with Resonant Hair}

For the Lagrangian in Eq. 2 of the main text, corresponding to a complex scalar field with a gauge coupling to 4-potential, the energy momentum $T_{\mu \nu} = T_{\mu \nu}^\text{EM} + T_{\mu \nu}^\Psi$ is
\begin{subequations}
\begin{align}
    \label{eq:Tmunu-EM}
    T_{\mu \nu}^\text{EM} &=
    F_\mu{}^{\alpha} F_{\nu \alpha} - \frac{1}{4} g_{\mu \nu} F_{\alpha \beta} F^{\alpha \beta} \,,
    \\
    \label{eq:Tmunu-Psi}
    T_{\mu \nu}^\Psi &=
    2\widetilde{\nabla}_{(\mu} \Psi^* \widetilde{\nabla}_{\nu)} \Psi +
    g_{\mu \nu} \left[ \widetilde{\nabla}^\alpha \Psi^* \widetilde{\nabla}_\alpha \Psi + V(\left| \Psi \right|) \right] \,,
\end{align}
\end{subequations}
where the scalar field potential is
\begin{equation}
    V(|\Psi|) = \mu^2 |\Psi|^2 (1 - 2 \lambda |\Psi|^2)^2 \,.
\end{equation}

We also rely on the 3+1 decomposition for BHs with resonant hair.
The electromagnetic four-potential is decomposed into components tangent and normal to the hypersurface,
\begin{equation}
    A_\mu = \mathcal{A}_\mu + A_n n_\mu \,,
\end{equation}
where $\mathcal{A}_\alpha \equiv \gamma_\alpha{}^\beta A_\beta$ and $A_n \equiv -n^\alpha A_\alpha$. The Maxwell tensor can then be written as
\begin{equation}
    \label{eq:Fmunu-decomposed}
    F_{\mu \nu} = {}^{(3)}D_\mu \mathcal{A}_\nu - {}^{(3)}D_\nu \mathcal{A}_\mu + n_\mu E_\nu - n_\nu E_\mu \,,
\end{equation}
where $E^\mu \equiv \gamma^{\mu \alpha} n^\beta F_{\alpha \beta}$ denotes the electric field.

By contracting the first index of (\ref{eq:Fmunu-decomposed}) with $n_\mu$ and projecting the second index with $\gamma^\nu{}_\beta$, we obtain the evolution equation for $\mathcal{A}^i$,
\begin{equation}
    \label{eq:EoM-Ai}
    \partial_t \mathcal{A}^i =
    - \alpha E^i
    - {}^{(3)}D^i(A_n \alpha )
    + \beta^j {}^{(3)}D_j \mathcal{A}^i - \mathcal{A}^j {}^{(3)}D_j \beta^i
    \,.
\end{equation}
Inserting (\ref{eq:Fmunu-decomposed}) into the equation of motion for $F_{\mu \nu}$, the normal projection yields the constraint
\begin{equation}
    \label{eq:EM-constraint}
    M \equiv {}^{(3)}D_i E^i - 4 \pi \rho_e = 0 \,,
\end{equation}
where $\rho_e \equiv - n_\mu j^\mu$ is the normal component of the four-current. The spatial projection provides the evolution equation
\begin{equation}
\begin{split}
    \label{eq:EoM-Ei}
   \partial_t E^i &=
        \alpha K E^i
        + \beta^j {}^{(3)}D_j E^i - E^j {}^{(3)}D_j \beta^i
        \\ &
        + {}^{(3)}D_j (\alpha ({}^{(3)}D^i \mathcal{A}^j - {}^{(3)}D^j \mathcal{A}^i) )
        - \alpha z^i
    \,,
\end{split}
\end{equation}
where $z^i \equiv \gamma^i{}_k j^k$ is the spatial projection of $j^\mu$. The $3+1$ decomposition of the current gives
\begin{subequations}
\begin{align}
    \rho_e &= - 2 i q (\Psi K_\Psi^* - \Psi^* K_\Psi)
              - 2 q^2 A_n \Psi \Psi^* \,,
    \\
    z^i &= - 2 i q \left( \Psi \partial^i \Psi^* - \Psi^* \partial^i \Psi \right)
           - 2 q^2 \mathcal{A}^i \Psi \Psi^*
    \,.
\end{align}
\end{subequations}
To fix the electromagnetic gauge freedom, we impose the Lorenz gauge condition, $\nabla_\mu A^\mu = 0$, which leads to the evolution equation for $A_n$,
\begin{equation}
    \label{eq:3+1-EoM-gauge}
    \partial_t A_n =
        \beta^i \partial_i A_n
        + \alpha K A_n
        - \alpha {}^{(3)}D_j \mathcal{A}^j - \mathcal{A}^j \partial_j \alpha
    \,.
\end{equation}

To suppress violations of the constraint (\ref{eq:EM-constraint}), we introduce an auxiliary variable, $Z$, and a constraint damping parameter, $\kappa$. The modified system is
\begin{subequations}
\begin{align}
    \begin{split}
        \label{eq:3+1-EoM-constraint-E}
        \partial_t E^i &=
            \alpha K E^i
            + \beta^j {}^{(3)}D_j E^i - E^j {}^{(3)}D_j \beta^i \\
            &+ {}^{(3)}D_j (\alpha ({}^{(3)}D^i \mathcal{A}^j - {}^{(3)}D^j \mathcal{A}^i) )
            + \alpha \partial^i Z
        - 4 \pi \alpha z^i
        \,,
    \end{split}
    \\
    \begin{split}
        \label{eq:3+1-EoM-constraint-gauge}
        \partial_t A_n &=
            \beta^i \partial_i A_n
            + \alpha K A_n \\
            &- \alpha {}^{(3)}D_j \mathcal{A}^j - \mathcal{A}^j \partial_j \alpha
            - \alpha Z
        \,.
    \end{split}
\end{align}
\end{subequations}
with the evolution equation for the auxiliary variable $Z$ given by
\begin{equation}
    \label{eq:3+1-EoM-constraint-Z}
    \partial_t Z =
        \alpha M
        - \alpha \kappa Z
        + \beta^j \partial_j Z
    \,.
\end{equation}

Using the first-order reduction for the scalar field defined above yields the same evolution equation for $\Psi$ (replace $\Phi$ by $\Psi$ in Eq. (\ref{eq:evo_Phi})), and substituting (\ref{eq:K_phi}) into the equation of motion yields the evolution equation for $K_\Psi$,
\begin{equation}
\begin{split}
    \label{eq:EoM-Kphi}
    \partial_t K_\Psi &=
    \alpha K K_\Psi +
    \beta^j \partial_j K_\Psi
    - \frac{1}{2} {}^{(3)}D^i(\alpha \partial_i \Psi)
    + \frac{1}{2} \alpha \frac{dV}{d\left| \Psi \right|^2} \Psi
    \\ &
    - i \alpha q \left( \mathcal{A}^i \partial_i \Psi - 2 K_\Psi A_n \right)
    + \frac{\alpha q^2}{2} \left( \mathcal{A}^i \mathcal{A}_i - A_n^2 \right) \Psi
    \,.
\end{split}
\end{equation}

\section{Appendix E - Convergence studies and constraint violations}
\label{apE}

We present here some convergence results that validate our numerical evolutions, by demonstrating that second-order convergence is obtained.
We first show convergence of the matter sector of the BHSH configuration \textbf{C} presented in the main text and Appendix A.
This is done by comparing the difference between various resolutions of the scalar field energy $E_\Phi$, and similarly of the angular momentum $J_{z,\Phi}$.
We then assess the agreement between $|E_2 - E_{1.6}|$ and $Q\times|E_{1.6} - E_{4/3}|$, where the subscript indicates the spacing of the coarsest grid from the corresponding simulation, and $Q$ is the expected convergence factor for second-order convergence, $Q=\frac{2^2-(1.6)^2}{(1.6)^2-(4/3)^2} \approx 1.84$.

We notice that the dynamics and, as a consequence, convergence properties, can be split into two distinct phases.
In Phase 1, the movement of the puncture is coming from numerical effects, and has small numerical values.
During this phase, from a physical point of view, the black hole is not properly moving, and the evolution of the scalar field is dominated by numerical loss.
In Phase 2, the position of the puncture reaches physically relevant values, and its movement is physical.
The interaction between the BH and the scalar field is the dominant effect and governs the dynamics.
However, we note that the starting point of the outspiral is resolution-dependent, resulting in marginal variations in the evolution, most notably a small time translation.
Therefore, to obtain a sensible comparison and before computing the differences mentioned above, we need to shift time series individually.
We do so by subtracting, in each simulation, a fiducial time $t_0$, chosen here to correspond to the moment when the puncture has a cylindrical radius $\rho\mu = 1$.
For the three simulations considered here, this evaluates to $t_0 \mu = 796.8, 715.52 \text{ and } 746.4$ respectively for spacings 2, 1.6 and 4/3.

\begin{figure}[tbph]
    \centering
    \includegraphics[width=0.75\linewidth]{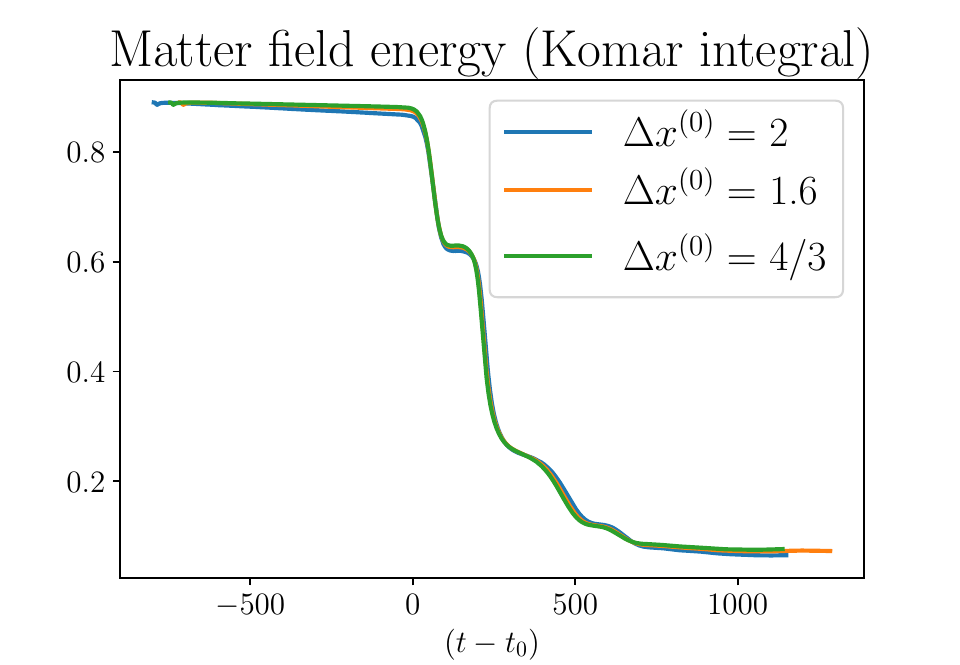}
    \caption{\small
    Evolution of the matter energy for BHSH configuration \textbf{C}, for three different resolutions, indicated by the spacing $\Delta x^{(0)}$ of the coarsest grid in the corresponding simulation.
    Time is shifted to make the respective puncture positions match, with $t_0$ corresponding to $\rho=1$ in each simulation.
    All quantities are given in units of $\mu$.
    }
    \label{fige1bis}
\end{figure}

\begin{figure*}[tbph]
    \centering
    \includegraphics[width=0.95\textwidth]{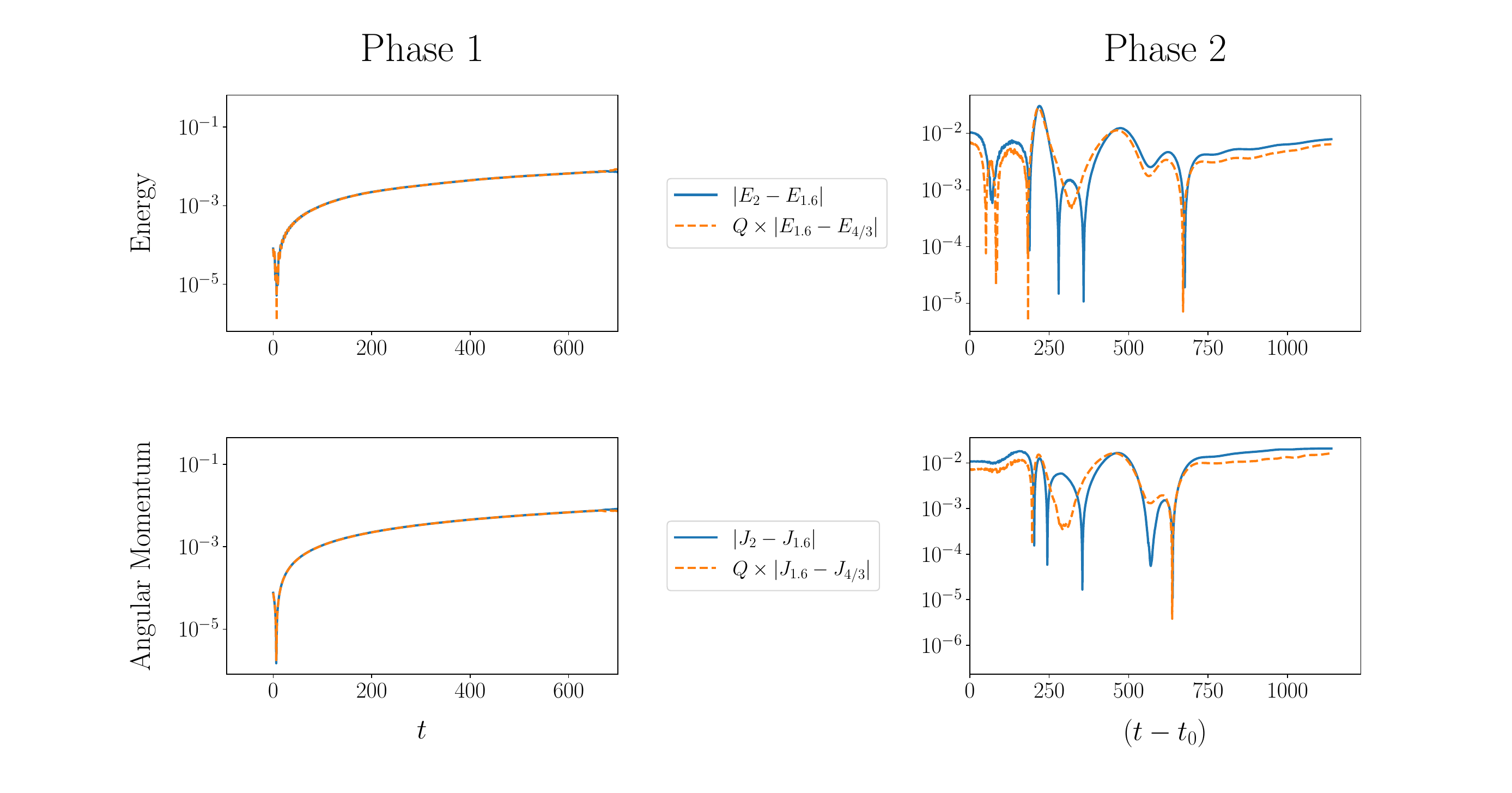}
    \caption{\small Second-order convergence of the matter energy $E_\Phi$ and angular momentum $J_{z,\Phi}$ for the BHSH configuration \textbf{C}, where the subscripts in the labels indicate the spacing of the coarsest grid from the corresponding simulation.
    See text for details.
    All quantities are given in units of $\mu$.
    }
    \label{fige1}
\end{figure*}

We exhibit in Fig.~\ref{fige1} the corresponding results.
The first row shows convergence of the scalar field energy, and the second row that of the angular momentum.
The first column shows results in the first phase, while the second column corresponds to Phase 2.
In the latter, the curves appear from $t=t_0$ onward, after proper individual time shifts as explained.
There is perfect superposition of the curves in Phase 1, when the errors are purely due to numerical loss.
Phase 2 also demonstrates very good agreement -- despite seeming discrepancies due to the differences changing sign and the logarithmic scale --, even ending with better-than-second-order convergence.
In addition, Fig.~\ref{fige1bis} shows the evolution of the matter energy for all three resolutions considered, after applying the respective time shifts.
The physicality and relevance of this approach appears clearly here.

\begin{figure*}[tbph]
    \centering
    \subfigure[\ BHsSH]{\includegraphics[width=0.4\textwidth]{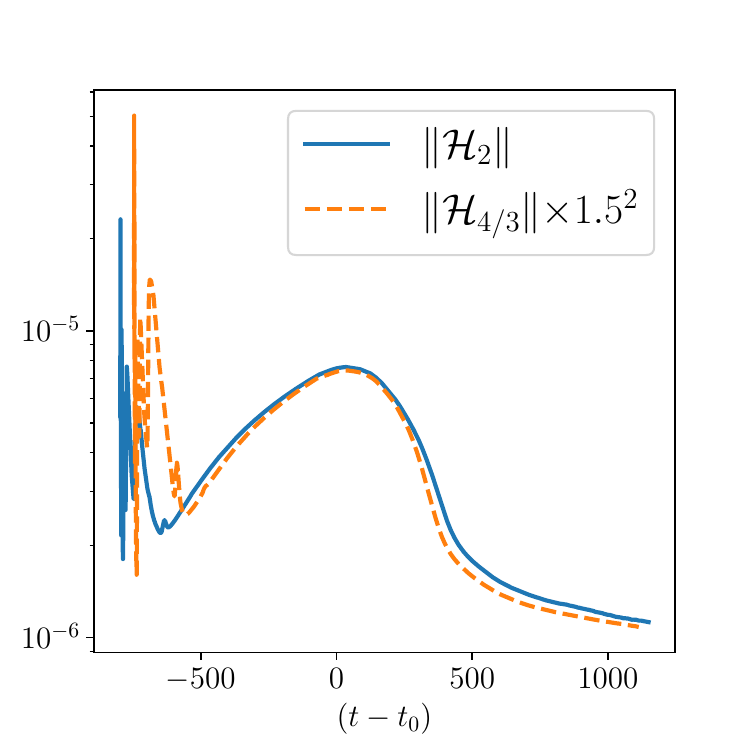}}
    \subfigure[\ BHsRH]{\includegraphics[width=0.4\textwidth]{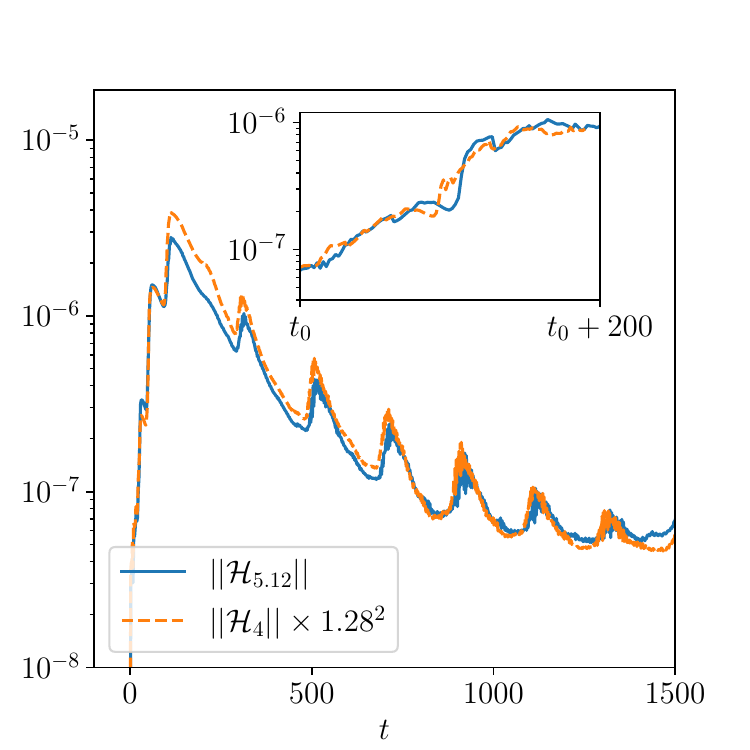}}
    \caption{Convergence to zero of the norm-2 of the Hamiltonian constraint $\mathcal{H}$.
    The subscript indicates the spacing of the coarsest grid from the corresponding evolution, and the higher resolution curve is rescaled by the proper factor to reflect second-order convergence.
    For the BHsSH (left), time is shifted to make the respective puncture positions match, with $t_0$ corresponding to $\rho=1$.
    Similarly, for the \mbox{BHsRH} (right), we show convergence in the first phase of the simulation, before any appreciable movement is seen (main panel), as well as the subsequent time-shifted curves, aligned at $t_0$ chosen so that the movements match (inset).
    All quantities are given in units of $\mu$.
    }
    \label{fige2}
\end{figure*}

We also exhibit in Fig.~\ref{fige2} the second-order convergence to zero of the Hamiltonian constraint, which must be satisfied in order to guarantee a physical solution of the BSSN system, both for a BHSH (configuration \textbf{C}) and a BHRH (upper branch of $r_H = 0.3$ and $\omega = 0.7$, in units of $\mu$).
In the case of the BHsSH (left), we perform the same time shift as before, to best demonstrate constraint convergence.
Similarly, for the BHsRH (right), the starting time of the BH ejection is resolution-dependent.
We show convergence in the first phase of the simulation, before any appreciable movement is seen (main panel), as well as the subsequent time-shifted curves, aligned at $t_0$ so that the movements match (inset).

\end{document}